\documentclass[twocolumn]{aastex631}

\usepackage{xspace}
\shorttitle{LIVIN' ON THE WEDGE}
\shortauthors{Balmer et al.}
\graphicspath{{./}{figures/}}
\definecolor{mylinkcolor}{HTML}{624CAB}
\definecolor{mylinkcolor2}{HTML}{8A2BE2}
\hypersetup{linkcolor=mylinkcolor,citecolor=mylinkcolor,urlcolor=mylinkcolor}

\newcommand{\um}{\text{\,\textmu m}\xspace}

\newcommand{\hra}{HR\,8799\xspace}

\newcommand{\hrb}{HR\,8799\,b\xspace}
\newcommand{\hrc}{HR\,8799\,c\xspace}
\newcommand{\hrd}{HR\,8799\,d\xspace}
\newcommand{\hre}{HR\,8799\,e\xspace}

\newcommand{\cE}{51\,Eri\xspace}
\newcommand{\cEb}{51\,Eri\,b\xspace}

\accepted{1 February 2025 to the Astronomical Journal}

\begin{document}

\title{JWST-TST High Contrast: Living on the Wedge, or,\protect\\NIRCam Bar Coronagraphy Reveals CO$_2$ in the \hra and \cE Exoplanets' Atmospheres}

\author[0000-0001-6396-8439]{William O. Balmer}
\correspondingauthor{William O. Balmer}
\email{wbalmer1@jhu.edu}
\affiliation{Department of Physics \& Astronomy, Johns Hopkins University, 3400 N. Charles Street, Baltimore, MD 21218, USA}
\affiliation{Space Telescope Science Institute, 3700 San Martin Drive, Baltimore, MD 21218, USA}

\author[0000-0003-2769-0438]{Jens Kammerer}
\affiliation{European Southern Observatory, Karl-Schwarzschild-Straße 2, 85748 Garching, Germany}

\author[0000-0003-3818-408X]{Laurent Pueyo}
\affiliation{Space Telescope Science Institute, 3700 San Martin Drive, Baltimore, MD 21218, USA}

\author[0000-0002-3191-8151]{Marshall D. Perrin}
\affiliation{Space Telescope Science Institute, 3700 San Martin Drive, Baltimore, MD 21218, USA}

\author[0000-0001-8627-0404]{Julien H. Girard}
\affiliation{Space Telescope Science Institute, 3700 San Martin Drive, Baltimore, MD 21218, USA}

\author[0000-0002-0834-6140]{Jarron M. Leisenring}
\affiliation{Steward Observatory and Department of Astronomy, University of Arizona, 933 N Cherry Ave, Tucson, AZ 85721, USA}

\author[0000-0002-6964-8732]{Kellen Lawson}
\affiliation{NASA-Goddard Space Flight Center, Greenbelt, MD 20771, USA}

\author[0009-0001-4688-6949]{Henry Dennen}
\affiliation{Department of Physics \& Astronomy, Amherst College, 25 East Drive, Amherst, MA 01002, USA}

\author[0000-0001-7827-7825]{Roeland P. van der Marel}
\affiliation{Space Telescope Science Institute, 3700 San Martin Drive, Baltimore, MD 21218, USA}
\affiliation{Department of Physics \& Astronomy, Johns Hopkins University, 3400 N. Charles Street, Baltimore, MD 21218, USA}

\author[0000-0002-5627-5471]{Charles A. Beichman}
\affil{NASA Exoplanet Science Institute, MS 100-22, California Institute of Technology, Pasadena, CA 91125, USA}

\author[0000-0001-5966-837X]{Geoffrey Bryden}
\affil{Jet Propulsion Laboratory, California Institute of Technology, 4800 Oak Grove Dr., Pasadena, CA 91109, USA}

\author[0000-0002-3414-784X]{Jorge Llop-Sayson}
\affil{Jet Propulsion Laboratory, California Institute of Technology, 4800 Oak Grove Dr., Pasadena, CA 91109, USA}

\author[0000-0003-3305-6281]{Jeff A. Valenti}
\affiliation{Space Telescope Science Institute, 3700 San Martin Drive, Baltimore, MD 21218, USA}

\author[0000-0003-3667-8633]{Joshua D. Lothringer}
\affiliation{Space Telescope Science Institute, 3700 San Martin Drive, Baltimore, MD 21218, USA}

\author[0000-0002-8507-1304]{Nikole K. Lewis}
\affiliation{Department of Astronomy and Carl Sagan Institute, Cornell University, 122 Sciences Drive, Ithaca, NY 14853, USA}

\author[0000-0002-2918-8479]{Mathilde Mâlin}
\affiliation{Department of Physics \& Astronomy, Johns Hopkins University, 3400 N. Charles Street, Baltimore, MD 21218, USA}
\affiliation{Space Telescope Science Institute, 3700 San Martin Drive, Baltimore, MD 21218, USA}

\author[0000-0002-4388-6417]{Isabel Rebollido}
\affiliation{European Space Agency (ESA), European Space Astronomy Centre (ESAC), Camino Bajo del Castillo s/n, 28692 Villanueva de la Ca\~nada, Madrid, Spain}

\author[0000-0003-4203-9715]{Emily Rickman}
\affiliation{European Space Agency (ESA), ESA Office, Space Telescope Science Institute, 3700 San Martin Dr, Baltimore, MD 21218, USA}

\author[0000-0002-9803-8255]{Kielan K. W. Hoch}
\affiliation{Space Telescope Science Institute, 3700 San Martin Drive, Baltimore, MD 21218, USA}

\author[0000-0003-2753-2819]{Rémi Soummer}
\affiliation{Space Telescope Science Institute, 3700 San Martin Drive, Baltimore, MD 21218, USA}

\author[0000-0003-4003-8348]{Mark Clampin}
\affiliation{Astrophysics Division, Science Mission Directorate, NASA Headquarters, 300 E Street SW, Washington, DC 20546, USA}

\author{C. Matt Mountain}
\affiliation{Association of Universities for Research in Astronomy, 1331 Pennsylvania Avenue NW Suite 1475, Washington, DC 20004, USA}

\begin{abstract}
High-contrast observations with JWST can reveal key composition and vertical mixing dependent absorption features in the spectra of directly imaged planets across the 3-5\um wavelength range. We present novel coronagraphic images of the \hra and 51~Eri planetary systems using the NIRCam Long Wavelength Bar (\texttt{LWB}) in an offset ``narrow" position. These observations have revealed the four known gas giant planets encircling \hra, even at spatial separations challenging for a 6.5\,m telescope in the mid-infrared, including the first ever detection of \hre at 4.6\um. The chosen filters constrain the strength of key molecules in each planet's atmosphere, notably CO$_2$ for the first time. The planets display a diversity of 3-5\um colors that could be due to differences in composition and ultimately be used to trace their formation history. They also show stronger CO$_2$ absorption than expected from solar metallicity models, indicating that they are metal enriched. We detected \cEb at 4.1\um and not at longer wavelengths, which, given the planet's temperature, is indicative of out-of-equilibrium carbon chemistry and an enhanced metallicity. Updated orbits fit to the new measurement of \cEb validate previous studies that find a preference for high eccentricities ($e{=}0.57_{-0.09}^{+0.03}$), which likely indicates some dynamical processing in the system's past. These results present an exciting opportunity to model the atmospheres and formation histories of these planets in more detail in the near future, and are complementary to future higher-resolution, continuum-subtracted JWST spectroscopy.
\end{abstract}


\section{Introduction}
\label{sec:introduction}

\subsection{Direct Imaging with JWST}

Direct imaging yields essential diagnostics of exoplanet atmospheric properties independent of the host star. Yet, given the formidable flux ratio (``contrast",  $C{\sim}10^{-3}{-}10^{-11}$) between stars and their planets, and the small angular separations ($\rho{\sim}1-1000\,\mathrm{mas}$) required for such measurements, the vast majority of imaged systems are young (${<}100\,\mathrm{Myr}$), with massive Jovian and super-Jovian planets on long-period orbits \citep[for an overview, see][]{Bowler2016, Pueyo2018, Currie2023, Follette2023}. Despite their paucity, directly imaged planets and their free-floating brown dwarf (BD) cousins are key to constraining the physical process underlying the formation, evolution, and atmospheric physics of giant planets \citep[e.g.,][]{Spiegel2012}. 

\par The widely separated, super-Jovian ($2{-}13\,M_{\rm J}$) population of directly imaged objects spans the low surface gravity L-T transition \citep[e.g.,][]{Faherty2016, Liu2016}, with $T_{\rm eff}{\sim}700{-}1500\,\mathrm{K}$, being detected with ground-based Adaptive Optics systems in the near-infrared ($1{-}4$\um). These observations probe orbits with $10{-}1000\,\mathrm{au}$, and these objects appear intrinsically rare \citep[e.g.,][]{Stone2018, Nielsen2019, Vigan2021} at these separations, more-so than giant planets orbiting at closer separations that have been detected indirectly \citep[e.g.,][]{Lagrange2023}. Due to its impressive sensitivity in the near- and mid-infrared, JWST \citep{Rigby2023} is sensitive to much cooler planets: either planets that are much older than known imaged planets \citep{Matthews2024} or young planets that are less massive \citep{Carter2021a}. Despite the improved sensitivity to cooler planets thanks to its wavelength grasp, JWST is not equipped with modern, deformable mirror assisted coronagraphs that will be key to the contrast performance of its successor, the Roman Space Telescope coronagraph \citep{Kasdin2020, Mennesson2020, Krist2023, Nemati2023}. As a result, in order to reach the deepest contrasts possible, JWST must rely on achieving moderate raw contrasts and gaining significant sensitivity during post-processing, using starlight subtraction algorithms and leveraging a combination of the telescope's exquisite stability \citep{Rigby2023, Telfer2024} and observational differential imaging strategies \citep{Soummer2014, Perrin2018, Girard2022, Greenbaum2023}.

\subsection{Coronagraphy with JWST/NIRCam}

\begin{figure*}
    \centering
    \includegraphics[width=0.9\textwidth]{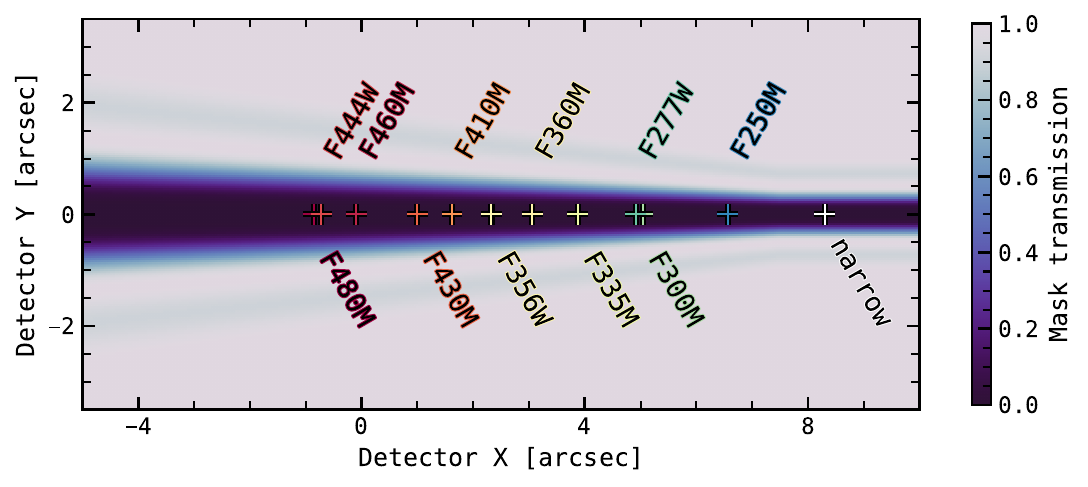}
    \caption{Model of the idealized two dimensional transmission function for the band-limited NIRCam MASKLWB coronagraphic mask in detector coordinates, generated using \texttt{webbpsf}. For each point in the detector frame, a filter-dependent offsets are indicated by crosses and labeled, including the ``narrow" offset position.}
    \label{fig:narrow_pos}
\end{figure*}

There are two sets of coronagraphs on NIRCam, the round masks and the wedge (or ``bar") masks \citep{Krist2009, Krist2010}. Both sets of coronagraph are Gaussian tapered occulters with apodized Lyot stops. These are effectively ``band-limited" coronagraphs \citep{Kuchner2002} that act on the intensity and not the phase of the incoming light, where the taper of the mask within a given bandwidth ($\epsilon D/\lambda)$ allows for transmission of sources close to the central obscuration, given an Lyot stop that rejects light at the edge of the pupil within that bandwidth. The Lyot stops for these coronagraphs were designed to be very robust to pupil shear and line-of-sight jitter, but reduce the effective aperture size of the instrument from 6.5\,m to 5.2\,m \citep{Mao2011}.\footnote{\url{https://jwst-docs.stsci.edu/jwst-near-infrared-camera/nircam-instrumentation/nircam-coronagraphic-occulting-masks-and-lyot-stops}}. The round masks were designed to provide symmetrical starlight suppression with a wide field of view; they cause 50\% loss in throughput at $6\lambda/D$ at 2.0\um (\texttt{MASK210R}, HWHM=0\farcs40), 3.5\um (\texttt{MASK335R}, HWHM=0\farcs64), and 4.3\um (\texttt{MASK430R}, HWHM=0\farcs82). The bar masks were designed to provide a preferentially smaller inner working angle, but sacrifice the field of view; they cause a 50\% loss in throughput at $4\lambda/D$ at filter-dependent positions; the short wavelength (SW) filters were designed to use the \texttt{MASKSWBAR} (HWHM=0\farcs13-0\farcs40), while the \texttt{MASKLWBAR} (HWHM=0\farcs29-0\farcs88) was designed to be used with the long wavelength (LW) channel filters (see Figure \ref{fig:narrow_pos}). Throughout this work, we refer to these coronagraphs without their `mask' prefix, e.g. the Long Wavelength Bar (\texttt{LWB}). In principle, the NIRCAM round masks are useful for unbiased searches for very faint planets at wide separations, while the bar masks are useful for efficient characterization of known planets at close separations.

Taking at face value the inner working angles (IWAs) of the NIRCam coronagraphs discussed above and projecting them at the distance of the typical members of young moving groups \citep[YMGs, e.g. $25{-}100\,\mathrm{pc}$,][]{Gagne2018, Gagne2024}, one might conclude that NIRCam coronagraphic observations cannot recover many of the currently known directly imaged giant planets. Moreover, since the peak of the semi-major axis distribution of giant planets is interior to the orbits of many directly imaged planets, around $10\,\mathrm{au}$ \citep[e.g.,][]{Fulton2021, Lagrange2023}, or $0\farcs05-0\farcs5$ projected separation for a typical YMG member, these geometric IWAs appear to preclude detecting a large fraction of fainter, yet undetected Jupiter mass exoplanets. However, the IWA is only the 50\% transmission point, and because of the observatory's impressive wave front stability \citep{Rigby2023}, sources interior to the nominal IWA have been detected with confidence as early as commissioning \citep{Girard2022, Kammerer2022}. 

Aside from studies during commissioning \citep{Kammerer2022, Girard2022} and GTO time \citep{Greenbaum2023}, the bar mask coronagraphs have been underutilized, likely owing to the relative complexity of observation planning given their preferential position angle (PA) requirements, their limited field of view, and the complexity of the treatment necessary for their associated data products. Throughout the first few years of JWST operations, the medium-sized round mask, \texttt{MASK335R}, has, on the other hand, seen extensive use. This is due to its balance between starlight suppression and IWA, its unbiased and wide field of view, and its relatively reliable target acquisition (TA) performance because of its sustained use. Thanks to the head-start this mode had during this period of intense community development and its position in the middle of trade-space, the Early Release Science (ERS) High-contrast community has recommended that observers proposing for point-source coronagraphic observations from 3-5\,\um opt for this mode for the time being \citep{Hinkley2023}. ERS results with the NIRCam round coronagraphs have found performance at or exceeding preflight expectations \citep{Carter2023}, achieving contrasts in the F444W filter of $10^{-5}$ at 1\farcs0. Recently, round mask observations of the AF~Leporis system detected one of the lowest-mass directly imaged planets, AF~Lep~b, at a separation of 0\farcs315 (or 1.8 $\lambda/D$, 7\% coronagraphic throughput) and placed upper limits on additional planets in the system down to sub-Saturn masses at {>}2\farcs0 \citep[][]{Franson2024}. A key question for potential observers, then, is where the inflection point occurs between starlight suppression performance (championed by the round masks) and source throughput (championed by the bar masks). Are there still science cases where it is better to trust post-processing techniques to suppress starlight and gather more photons from a planet, despite a limited field of view, using the NIRCam bar masks?

\par Characterization efforts during commissioning suggest that the answer is yes. The bar mask has performed about a magnitude better in contrast at closer separations ($<0\farcs75$) than the round mask in an equivalent filter, while the opposite is true at wider separations (\citealt{Girard2022}, and see Figure 8 in \citealt{Kammerer2022}). The obvious advantage of increased throughput is lower integration time and better performance for filters with smaller bandpasses. The impressive close-in detections with the round masks noted above were taken in the widest filters available on NIRCam, the F444W or F356W filters, which cover multiple absorption features and are therefore less sensitive to the particular composition of the planet than the medium or narrowband filters. As noted in \citet{Perrin2018}, there is untapped contrast performance at close separations that can be gained with the wedge masks simply by ignoring the filter-dependent positions along the bar. Both the SW and LW bar masks aboard NIRCam taper toward a ``narrow" end, and there now exists an engineering offset position behind this part of both masks (see Figure \ref{fig:narrow_pos}). The addition of this offset position was directly inspired by the successful in-flight commissioning of HST/STIS's \texttt{BAR5} coronagraphic mask \citep{debes2017chasing}. This is a ``shared-risk" observation strategy, being an offered but yet-unsupported mode. \citet{Perrin2018} predicted that observations using this ``narrow" offset position would reach the requisite contrasts to detect the lowest-mass (and one of the closest separation) directly imaged planets known at the time, \cEb \citep{Macintosh2015}, as well as the innermost planets in the quintessential direct imaging target system, \hra \citep{Marois2008} in multiple medium filters, enabling the characterization of their 3-5\um spectral shape.

\subsection{This Paper}
\par To this end, in this paper we characterize the performance of the NIRCam \texttt{LWB} coronagraph's narrow offset, revealing its ability to efficiently characterize giant planets at blisteringly close separations ($3{-}10\,\mathrm{pixels}\simeq0\farcs25{-}0\farcs5$). Despite the glut of observations of our two target systems in the near infrared, \hre remains undetected from the ground at 4.6\um in the \textit{M} band \citep{Galicher2011}, and the 3-5\um spectral shape of these planets remains uncertain, especially between 4-5\um where telluric absorption is strong \citep{Skemer2014, Doelman2022}. These wavelength ranges carry crucial information about the physics and chemistry of these atmospheres, which must be solved in order to truly understand their underlying composition.  
\par This paper is part of a series by the JWST Telescope Scientist Team (JWST-TST)\footnote{\url{https://www.stsci.edu/~marel/jwsttelsciteam.html}}. This collaboration uses Guaranteed Time Observations (GTO, PI: M.~Mountain) for projects across three topic areas: Exoplanet and Debris Disk High-Contrast Imaging (lead: M.~Perrin), Transiting Exoplanet Spectroscopy (lead: N.~Lewis), and Local Group proper-motion Science (lead: R.~van der Marel). Previous work from the TST High-contrast series includes \citet{Rebollido2024, Kammerer2024, Ruffio2024, Hoch2024}. 
\par In \S\ref{sec:targets}, we introduce the two directly imaged planetary systems that we targeted with the NIRCam \texttt{LWB}. The observations and data reduction are described in \S\ref{sec:data}. In \S\ref{sec:analysis} we analyze the 3-5\um colors of the \hra planets by comparing them to previous observations and an empirical sample; we also fit atmospheric models and orbits to our observation of \cEb. \S\ref{sec:discussion} discusses our results, and we conclude in \S\ref{sec:conclusion}.

\section{Target Systems} \label{sec:targets}

\subsection{HR~8799}

\par The \object[HR 8799]{HR\,8799} system is a directly imaged system of four self-luminous giant planets in orbit around a young, chemically peculiar, debris disk hosting A-type star. The system presents a number of challenges to our understanding of planet formation and evolution, and has remained emblematic since its discovery. The photometric variability and chemical peculiarity of \hra~A was noted during the late 20th century \citep{Gehren1977, Schuster1986, Rodriguez1995}, and the star was subsequently classified as an A-type $\lambda$~Bootis (meaning sub-solar iron-peak abundances, but near-solar C, N, O, S abundances), $\gamma$~Dor variable \citep{Zerbi1999, Gray1999, Moya2010, Wright2011, Sodor2014, WangJi2020, Sepulveda2023, Hubrig2023}. The system's parallax as determined by the Gaia spacecraft is $\pi=24.462{\pm}0.0455\,\mathrm{mas}$ \citep{GaiaCollaboration2022}, which corresponds to a distance of $40.80{\pm}0.08\,\mathrm{pc}$. \hra~A's proper-motion makes it a probable, though somewhat isolated, member\footnote{See \citet[Appendix A]{Faramaz2021} for an excellent overview of the system's apparent isolation and possible young moving group associations. In short, it appears that the star formed in relative isolation nearby the protomembers of the Columba and Carina associations, about $30-40\,\mathrm{Myr}$ ago.} of the $\sim40\,\mathrm{Myr}$ Columba association \citep[e.g.,][]{Doyon2010, Zuckerman2011, Bell2015, Gagne2018}. The star's radius has been measured with optical interferometry ($R_\star = 1.44\pm0.06\,\mathrm{R_\odot}$), which provides a precise age and mass constraint when coupled with literature photometry and distance measurements \citep[$33^{+7}_{-13}\,\mathrm{Myr}$ and $M_\star=1.516^{+0.038}_{-0.024}\,\mathrm{M_\odot}$,][]{Baines2012}. In addition, the star's mass has been estimated using the Keplerian orbits of the 4 known giant planets \citep[$1.43^{+0.06}_{-0.07}\,\mathrm{M_\odot}$,][]{Sepulveda2022a}. While aspects of the stellar pulsations, photospheric abundances, and kinematic association remain a mystery, overall, the basic stellar properties are now well constrained and help inform our measurements of the planets.
\par The presence of debris encircling the star was also identified in the late 20th century, by observations of excess emission at $60\um$ \citep{Sadakane1986}. The system's debris belt was subsequently characterized with mid- and far-infrared imaging and submillimeter interferometry, revealing multiple components: an inner warm ($\sim150~\mathrm{K}$) dust belt from $5-15$ au, and an outer cold ($\sim50~\mathrm{K}$) dust belt from $90-300$ au \citep{Chen2006, Su2009, Chen2009, Moro-Martin2010, Hughes2011, Matthews2014, Booth2016, Geiler2019, Faramaz2021}. Most recently, emission from the inner debris belt was imaged with JWST/MIRI at 15\um \citep{Boccaletti2023}.

\par In 2008, \citet{Marois2008} announced the detection of three objects, apparently of planetary mass given their near infrared colors and apparent magnitudes, that shared common proper-motion with and appeared to orbit \hra\,A. They were situated between 15 and 80 au, and became known as \hrb, \hrc, and \hrd. These planets were some of the first to be directly imaged \citep{Chauvin2004, Lagrange2009}, and the multiplanet nature of the discovery distinguished the system from other direct imaging discoveries at the time. In 2010, \citet{Marois2010} announced the detection of a fourth planet, interior to the three, \hre. Subsequent studies detected various combinations of the four planets in a variety of archival observations dating as far back as 1998 \citep{Lafreniere2009}, and new observations have continued to monitor the system in order to measure, with increased precision, the orbital motion and atmospheric properties of the planets \citep[e.g.,][]{Wang2018, Sepulveda2022a, Zurlo2022, Thompson2023}.
\par Spectral and photometric characterization of these planets has a storied history, tracing the development of many high-contrast methods \citep[e.g.,][]{Fukagawa2009, Bowler2010, Galicher2011, Soummer2011, Konopacky2013, Bonnefoy2016, Greenbaum2018, Ruffio2019, GRAVITYCollaboration2019, PetitditdelaRoche2020, Biller2021, Wang2022} and substellar atmosphere models \citep[e.g.,][]{Barman2011, Marley2012, Barman2015, Molliere2020, Wang2023}. Observations from JWST/MIRI and NIRCam have detected all four planets from space, placing strong constraints on their bolometric luminosity \citep{Boccaletti2023, Bryden2024_placeholder}. The architecture of the system---two debris belts sandwiching four gaseous planets---is reminiscent of a ``scaled-up" version of our own solar system. Most recently, \citet{Nasedkin2024} presented an exhaustive atmospheric analysis of the planets based on new VLTI/GRAVITY observations that indicated a preference for very metal enriched atmospheres and non-equilibrium carbon and oxygen chemistry driven by vertical mixing. They indicated that the 3-5\um wavelength range is especially sensitive to differences between certain classes of their atmospheric models (see their Figures 6 and 19) that are driven by changes in the vertical mixing parameter, carbon-to-oxygen (C/O) ratio, or the atmospheric metallicity ([Fe/H]).
\par Close cousin to this paper, \citet{Bryden2024_placeholder} as part of the joint NIRCam/MIRI/TST GTO 1194 (PI: C. Beichman), have imaged the \hra system with JWST/NIRCam using the \texttt{MASK335R} coronagraph and the wide F356W and F444W filters. This program was designed to search the outer regions of the system (between \hrb and the debris ring) for additional planets. Our observations, complementary to theirs, probe the known planetary system to better characterize their composition.

\subsection{51~Eri}

\par The Jovian-mass planet \cEb is the jewel in the crown of the Gemini Planet Imager Exoplanet Survey: a young planet with strong signatures of methane in its atmosphere on a 10\,au orbit \citep{Macintosh2015}. Unlike wider separation, comoving companion systems that are less amenable to age estimation or orbital dynamical mass constraints, like GJ~504~b \citep{Kuzuhara2013} or COCONUTS-2~b \citep{Zhang2021}, \cEb has a firmly planetary mass. Dynamical mass constraints on the orbit give $<11\,M_{\rm J}$ at $2\,\sigma$ confidence, because of the effective lack of a proper-motion anomaly measured between Hipparcos and Gaia \citep{DeRosa2020, Dupuy2022}. The host star, \cE, is an F0-type member of the 24~Myr $\beta$-Pic moving group (BPMG), so the system has a well-defined association, even if the exact age for the BPMG is still debated \citep[e.g.,][]{Bell2015, Lee2024}, and so the luminosity of the planet constrains its mass to be less than $\lesssim12\,M_{\rm J}$, given the planet's uncertain initial entropy \citep{Macintosh2015}. The radius of \cE has been measured with optical interferometry, and so its mass and age have been constrained isochronally \citep[$R_\star=1.45\pm0.02\,\mathrm{R_\sun}$, $M_\star=1.550\pm0.006\,\mathrm{M_\sun}$, Age$=23.2^{+1.7}_{-2.0}\,\mathrm{Myr}$,][]{Elliott2024}. The star is also a $\gamma$-Doradus type pulsator, and further characterization of its pulsation modes could contribute to a better understanding of its properties \citep{Sepulveda2022b}. The system hosts a cold ($50\,\mathrm{K}$) debris disk with an inner edge at $\sim80\,\mathrm{au}$ (exterior to \cEb), as evidenced by excess in the far-infrared observed by \textit{Herschel}. This debris has a much lower fractional luminosity, $L_{\rm IR}/L_\star=2.0\times10^{-6}$, than other members of the $\beta$-Pic moving group \citep{Riviere-Marichalar2014}.
\par \cEb has been observed in the near infrared from the ground, but due to its intrinsic faintness and high-contrast, has been subjected to less scrutiny than the \hra planets. The available data indicates a high-eccentricity orbit \citep{Maire2019, DeRosa2020, Dupuy2022}, but this estimate has been uncertain due to limited orbital coverage. Its atmosphere appears to be cool ($\sim700-800\,\mathrm{K}$), at least partially cloudy, with the potential for thin haze layers to form in the upper atmosphere, and exhibits some signs of chemical disequilibrium driven by vertical mixing \citep{Zahnle2016, Rajan2017, Samland2017, Tsai2021, Brown-Sevilla2023, Madurowicz2023, Whiteford2023}. Like the \hra planets, observations between the \textit{L'} and \textit{M} band are necessary to constrain the strength of vertical mixing and cloud opacity in the atmosphere \citep[see discussion in][]{Rajan2017, Madurowicz2023}.

\section{Observations and Data Reduction}
\label{sec:data}

\subsection{Observing Strategy} \label{subsec:observing_strategy}

These observations were taken as a part of the GTO programs 1412 (PI: Perrin) on 2023 October 18\footnote{These observations were initially attempted on 2022 October 20, but failed due to a coronagraph centering error and were rescheduled.} and 1194 (PI: Beichman) on 2023 November 5. The observations used the \texttt{MASKLWB} and the ``narrow" fiducial point override, targeting medium-band filters between $3{-}5\um$ where ground-based data was sparse or absent; simultaneous SW filters were observed due to the NIRCam dichroic. Observations of the two associated reference stars used the \texttt{5-POINT-BAR}\footnote{\url{https://jwst-docs.stsci.edu/jwst-near-infrared-camera/nircam-operations/nircam-dithers-and-mosaics/nircam-subpixel-dithers/nircam-small-grid-dithers\#NIRCamSmallGridDithers-Coronagraphicimaging}} small grid dither pattern to improve the diversity of the sampled point-spread-functions \citep[PSFs;][]{Soummer2014, Lajoie2016} The observations, including filters, detector readout parameters, exposure times, and telescope PAs are recorded in Table \ref{tab:observations}. 
\par
\begin{figure*}
\gridline{\fig{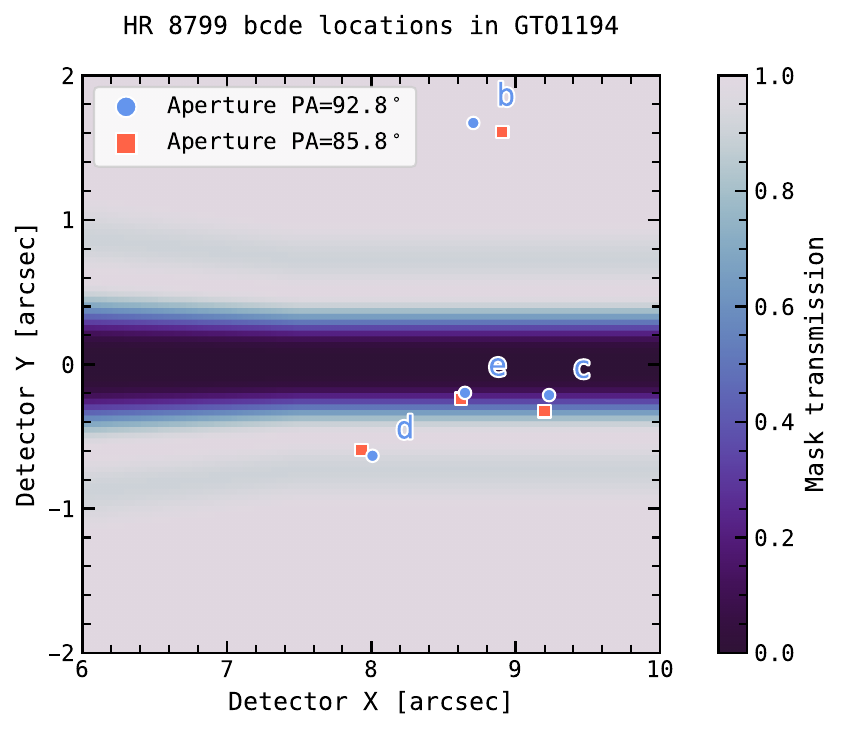}{0.45\textwidth}{}
          \fig{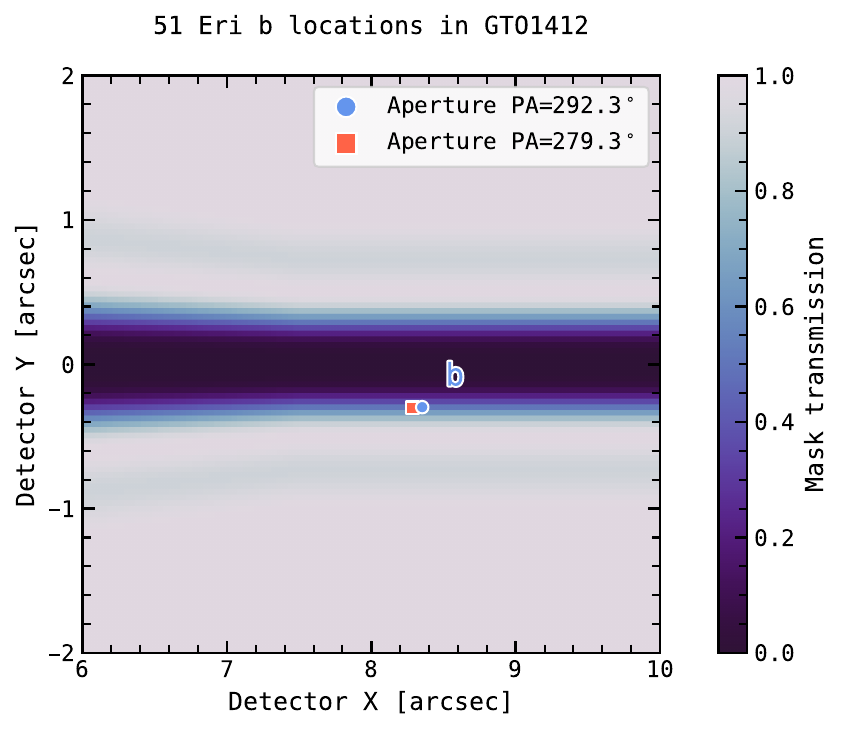}{0.45\textwidth}{}
          }
% two column
% \gridline{\fig{figures/HR8799_planetpos_1194}{0.4\textwidth}{}}
% \gridline{
%           \fig{figures/51Eri_planetpos_1412}{0.4\textwidth}{}
%           }
\caption{Planet locations per roll angle in detector coordinates at the epoch of observation, compared to the idealized coronagraph transmission function of the NIRCam \texttt{LWB}/narrow offset position. \textit{Left}: Illustrating the position of the four \hra planets in program 1194. \textit{Right:} Illustrating the position of \cEb in program 1412. The first roll angle position is indicated with blue circles, the second with red squares, and the legend indicates the position angle of the telescope aperture for each roll. \textit{Takeaway:} This figure also explains the difference in our forward modeling analysis for each system. In the \cE observations, the coronagraphic throughput at the location of the faint planet does not change significantly between roll angles, and we opt to forward model the planet simultaneously in both roll angles using an \texttt{ADI+RDI} starlight subtraction. In the \hra observations, the coronagraphic throughput at the location of the relatively brighter planets changes significantly between roll angles, and \hre suffers from strong self-subtraction due to the orientation of the position-dependent PSF and the roll angle, motivating a roll by roll forward modeling analysis with a \texttt{RDI} starlight subtraction. } \label{fig:positions_mask}
\end{figure*}

 The positioning of the coronagraph requires PA restrictions based on the predicted location of the planets based on prior orbit monitoring, to ensure the most optimal throughput at the location of the target. We used initial guesses of each \hra planet's position based on unpublished, but publically accessible\footnote{\url{https://www.whereistheplanet.com/}} orbits \citep[A. Chavez priv. comm.,][]{whereistheplanet}. Our predictions for the location of \cEb were based on our own reproduction of the best-fitting orbits in \citet{Dupuy2022} using \texttt{orbitize!} \citep[][and see \S\ref{subsec:51eri_analysis} for a more complete description]{Blunt2020}. Figure \ref{fig:positions_mask} indicates the positions of the planets in each roll angle with respect to the coronagraph transmission (centered at the ``narrow" offset position; Figure \ref{fig:narrow_pos}). These predictions are also used for forward modeling our detections in the post-starlight-subtracted images.

\begin{table*}[!t]
    \footnotesize
    \centering
    \caption{GTO 1194 and GTO 1412 NIRCam \texttt{LWB}/narrow Observing Log.}
\label{tab:observations}
    \begin{tabular}{c c c c c c c c c}
        Obs. ID & Target & SW Filter & LW Filter & Readout pattern & Dither pattern & $N_{\rm{ints}}$/$N_{\rm{groups}}$/$N_{\rm{frames}}$ & $T_{\rm{exp}}$ [s] & PA$^\dag$ [deg] \\
        \hline
        \hline
        1194, 2 & \hra & F200W & F250M & BRIGHT2 & NONE & 5 /115/115 & 1347.0   & 92.8 \\
        1194, 2 & \hra & F200W & F300M & BRIGHT2 & NONE & 8 / 75/75  & 1356.81  & 92.8 \\ 
        1194, 2 & \hra & F182M & F335M & BRIGHT2 & NONE & 10/ 60/60  & 1340.558 & 92.8 \\
        1194, 2 & \hra & F182M & F410M & BRIGHT2 & NONE & 10/ 30/30  & 670.279  & 92.8 \\
        1194, 2 & \hra & F210M & F430M & BRIGHT2 & NONE & 10/ 60/60  & 1340.558 & 92.8 \\
        1194, 2 & \hra & F210M & F460M & BRIGHT2 & NONE & 10/ 40/40  & 893.706  & 92.8 \\
        \hline
        1194, 4 & HD~220657 & F200W & F250M & BRIGHT2 & 5-POINT-BAR & 4/24/120 & 1150.454 & 90.4 \\
        1194, 4 & HD~220657 & F200W & F300M & BRIGHT2 & 5-POINT-BAR & 4/24/120 & 1150.454 & 90.4 \\
        1194, 4 & HD~220657 & F182M & F335M & BRIGHT2 & 5-POINT-BAR & 4/24/120 & 1150.454 & 90.4 \\
        1194, 4 & HD~220657 & F182M & F410M & BRIGHT2 & 5-POINT-BAR & 4/14/70  & 671.098  & 90.4 \\
        1194, 4 & HD~220657 & F210M & F430M & BRIGHT2 & 5-POINT-BAR & 4/24/120 & 1150.454 & 90.4 \\
        1194, 4 & HD~220657 & F210M & F460M & BRIGHT2 & 5-POINT-BAR & 4/18/90  & 862.841  & 90.4 \\
        \hline 
        1194, 5 & \hra & F200W & F250M & BRIGHT2 & NONE & 5 /115/115 & 1347.0   & 85.8 \\
        1194, 5 & \hra & F200W & F300M & BRIGHT2 & NONE & 8 / 75/75  & 1356.81  & 85.8 \\ 
        1194, 5 & \hra & F182M & F335M & BRIGHT2 & NONE & 10/ 60/60  & 1340.558 & 85.8 \\
        1194, 5 & \hra & F182M & F410M & BRIGHT2 & NONE & 10/ 30/30  & 670.279  & 85.8 \\
        1194, 5 & \hra & F210M & F430M & BRIGHT2 & NONE & 10/ 60/60  & 1340.558 & 85.8 \\
        1194, 5 & \hra & F210M & F460M & BRIGHT2 & NONE & 10/ 40/40  & 893.706  & 85.8 \\
        \hline
        1412, 12 & \cE & F182M & F335M & BRIGHT2 & NONE & 10/40/40 & 893.706  & 292.3 \\
        1412, 12 & \cE & F182M & F410M & BRIGHT2 & NONE & 10/40/40 & 893.706  & 292.3 \\
        1412, 12 & \cE & F210M & F430M & BRIGHT2 & NONE & 10/60/60 & 1340.558 & 292.3 \\
        1412, 12 & \cE & F200W & F460M & BRIGHT2 & NONE & 10/40/40 & 893.706  & 292.3 \\
        \hline
        1412, 14 & HD~30562 & F182M & F335M & BRIGHT2 & 5-POINT-BAR & 10/20/100 & 2234.264  & 287.5 \\
        1412, 14 & HD~30562 & F182M & F410M & BRIGHT2 & 5-POINT-BAR & 10/18/90  & 2010.838  & 287.5 \\
        1412, 14 & HD~30562 & F210M & F430M & BRIGHT2 & 5-POINT-BAR & 10/28/140 & 3127.97   & 287.5 \\
        1412, 14 & HD~30562 & F200W & F460M & BRIGHT2 & 5-POINT-BAR & 10/18/90  & 2010.838  & 287.5 \\
        \hline
        1412, 15 & \cE & F182M & F335M & BRIGHT2 & NONE & 10/40/40 & 893.706  & 279.3 \\
        1412, 15 & \cE & F182M & F410M & BRIGHT2 & NONE & 10/40/40 & 893.706  & 279.3 \\
        1412, 15 & \cE & F210M & F430M & BRIGHT2 & NONE & 10/60/60 & 1340.558 & 279.3 \\
        1412, 15 & \cE & F200W & F460M & BRIGHT2 & NONE & 10/40/40 & 893.706  & 279.3 \\
        \hline
    \end{tabular}
    \tablecomments{$\dag$ The angle recorded here is the ``Aperture PA" angle. For the NIRCam coronagraphic mode, there is an offset between this value and the PA of the telescope's V3 axis. Both PA\_V3 and the Aperture PA are recorded in the image \textit{.fits} file headers.}
\end{table*}

\subsection{Image processing} \label{subsec:image-processing}

\par We reduced our data using the python package \texttt{spaceKLIP}, following previous work \citep{Kammerer2022, Carter2023, Franson2024, Lawson2024}. \texttt{spaceKLIP} is a community-developed code that provides a user-friendly interface for many \texttt{jwst} pipeline \citep{Bushouse2023} steps with modifications appropriate for coronagraphic imaging, and creates inputs for the \texttt{pyklip} package \citep{Wang2015} that is used to perform starlight subtraction (see \S\ref{subsec:psfsub} below). Since data reduction using \texttt{spaceKLIP} has been described exhaustively elsewhere, below we described the specific changes we implemented to treat the \texttt{LWB}/narrow observations, and we refer the reader to previous work \citep{Kammerer2022, Carter2023, Franson2024, Kammerer2024} as well as the \texttt{spaceKLIP} documentation\footnote{\url{https://spaceklip.readthedocs.io/en/latest/}} for additional details. The reduction presented here used the \texttt{jwst} pipeline version 1.12.1 and the Calibration Reference Data System (CRDS) version 11.17.19, with the associated \texttt{jwst\_1256.pmap} file, and the associated flux calibrations/filter zero-points for our measurements of the apparent magnitudes and flux in microjanskys for each planet. For the relative flux measurements (contrast, $\Delta\mathrm{mag}$) we reference the filters and their associated zero-points from the Spanish Virtual Observatory Filter Profile Service \citep[SVO FPS;\footnote{\url{http://svo2.cab.inta-csic.es/svo/theory/fps/index.php}}][]{Rodrigo2020} and use the latest calibrated spectrum of Vega from CALSPEC\footnote{\url{https://www.stsci.edu/hst/instrumentation/reference-data-for-calibration-and-tools/astronomical-catalogs/calspec}} \citep{Bohlin2014, Bohlin2020}. 
\par In short, we reduced the raw (Stage 0, ``uncal.fits") files into flux-calibrated (Stage 2, ``calints.fits") files using the \texttt{jwst} pipeline, ignoring the NIRCam dark subtraction step due to low signal-to-noise-ratio (SNR) reference files and selecting 4 pixels along each edge of the subarray to use as ``pseudo-reference" pixels for the jump correction step (with a jump threshold of 4). We corrected for bad pixels not flagged by the \texttt{jwst} pipeline using sigma-clipping with a threshold of 5. Due to the undersampled PSF, the images are then blurred above the Nyquist sampling threshold. The images are blurred using a Gaussian kernel with full-width-at-half-maximum of $\mathrm{FWHM} = \lambda/2.3D$, where $\lambda$ is the central wavelength of the given filter, and $D=5.2$m to account for the effective aperture of the JWST/NIRCam Lyot stops. This blurring step is important to conduct before any interpolations (such as shifting the image) are conducted, to avoid Fourier interpolation artifacts. We padded/cropped the NIRCam bar 400x256 pixel subarray to 451x451 pixels. We then translate, subtract, and minimize the residuals between the first science frame and a perfectly centered (both within the image array, and behind the coronagraphic mask) \texttt{webbpsf} \citep{Perrin2012, Perrin2014} model of the coronagraphic PSF generated using \texttt{webbpsf\_ext}\footnote{\url{https://github.com/JarronL/webbpsf_ext}} in order to determine a centering offset. This model, like all other \texttt{webbpsf} models mentioned in this work, was generated using the observatory's nearest-in-time wavefront sensing measurement as input for the Optical Path Difference\footnote{See, for instance, \url{https://webbpsf.readthedocs.io/en/latest/jwst_measured_opds.html}, or the simulations in \citet{Perrin2018}.}. We translate, subtract, and minimize the residuals between the remaining frames and the centered first science frame to determine the centering (for science frames) and small grid dither (for reference frames) offsets, and Fourier shift the images by these measured offsets to register and center the entire dataset. In this process, the intention is to align the rolls and dithers (which have small subpixel displacements with respect to one-another due to imperfect centering behind the coronagraphic mask or due intentionally to the small grid dither pattern) to the center of the image array, so that any future rotations are symmetric about the image and any future decompositions of the PSF are aligned. Instead of conducting PSF subtraction with the unaligned library of images, centering the entire sequence in this way allows each instance of the PSF to contribute to the principle component analysis of the target PSF. We then cropped the 451x451 pixel images down to 201x201 pixels (3\farcs2 in LW and 6\farcs5 in SW), the region of astrophysical interest.
\par Due to a \textit{.fits} writing error in the creation of the \texttt{LWB}/narrow 400x256 pixel subarrays, the \*uncal.fits files available on MAST as of 2024 June 1st do not have correct CRPIX values in their headers, which means that the position of the coronagraphic mask (and therefore the throughput) cannot be determined by referencing the header coordinates for a given image (in this specific mode and sub-array). We therefore implemented a small change in \texttt{spaceKLIP} that allows the user to optionally update the headers for images with the appropriate values from the Science Instrument Aperture Files (SIAF) using the \texttt{pySIAF}\footnote{\url{https://pysiaf.readthedocs.io/en/latest/}} package. 

\begin{figure*}
    \centering
    \includegraphics[width=0.9\textwidth]{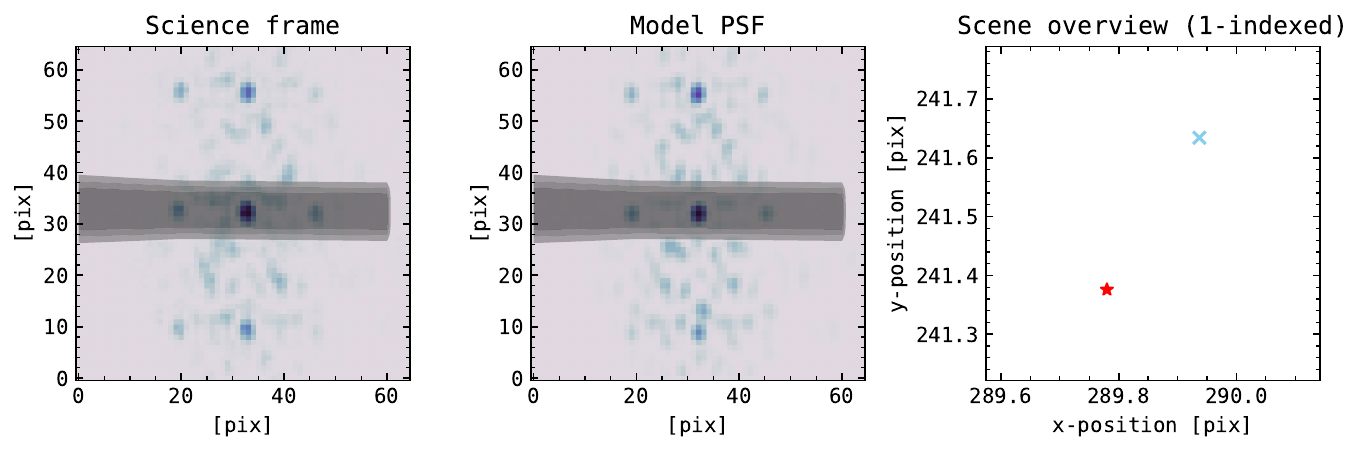}
    \caption{Centering NIRCam \texttt{LWB}/narrow coronagraphic images. \textit{Left}: 1st science frame from the \hra observations in the F460M filter (the filter with the farthest distance between its typical filter-dependent offset and the narrow end of the wedge), in arbitrary pixel coordinates. The central leakage term is roughly centered, and the PSF wings appear about 20 pixels above and below this leakage; the field of diffraction peaks due to the edges of the Lyot stop are visible to either side of the bar, and are sensitive to the position of the star behind the coronagraph. \textit{Middle}: \texttt{webbpsf} model PSF based on the nearest-in-time OPD observations, in arbitrary pixel coordinates. \textit{Right}: Overview of relevant positions in the padded 451x451 pixel coordinates; the mask center for this filter is indicated by a blue cross and the computed star position behind the mask by a red star.}
    \label{fig:centering}
\end{figure*}

\par We also found that the default method in \texttt{spaceKLIP} for registering coronagraphic images did not recover shifts consistent with the commanded small grid dithers for our  \texttt{LWB}/narrow observations, which resulted in an improperly centered image library. The cross correlation between the central leakage term (the peak of the diffraction of the longer wavelength PSFs directly behind the coronagraph mask) at one end of the \texttt{5-POINT-BAR} dither pattern and the other end does not appear to be a robust measure of the underlying shift at a subpixel level (a distance of about 40\,mas). The cross correlation of the off-axis diffraction pattern of the PSF proved much more sensitive to the position of the star behind the mask. We implemented two optional tools to down-weight the leakage terms and up-weight the diffraction pattern. We implemented a method for masking the central leakage term with a rectangle of zero values during the shift estimation step. We also found that taking the square root of the image before computing the cross correlation improved the accuracy of our subpixel shifts. Combining these two strategies, we were able to successfully recover the expected difference in position between the reference star images along the dither pattern. Taking the square root of the images de-emphasizes the two bright ``wings" of the PSF in comparison to the fainter, off-axis diffraction peaks. This square root is applied only during the shift computation, not to the shifted images themselves. Figure \ref{fig:centering} shows an example NIRCam \texttt{LWB}/narrow PSF in the F460M filter, the \texttt{webbpsf} model used for centering, and an overview of the relevant detector coordinates.

\subsection{Starlight subtraction and Point-source Forward Modeling} \label{subsec:psfsub}

\par We subtracted the residual starlight from our reduced images using the Karhunen--Lo\`{e}ve Image Projection (KLIP) algorithm \citep{Soummer2012}. \texttt{spaceKLIP} wraps the package \texttt{pyklip}\footnote{\url{https://pyklip.readthedocs.io/en/latest/}} \citep{Wang2015} that is a python implementation of the KLIP algorithm. KLIP enables ``forward modeling" of astrophysical sources through the starlight subtraction process so that one can estimate the malign affects of the algorithm on the throughput and morphology of the planet's PSFs, and account for algorithmic degradation when measuring the astrometry and photometry \citep{Pueyo2016}. The model of the residual starlight is determined from a reference library composed of the other images. The best-fitting model that describes the entire speckle field may perform worse in specific regions of the speckle pattern, especially when there are mismatches between the PSF of the target and reference observations that can result from changes in the centering behind the coronagraph, dissimilar stellar spectral types or magnitudes, and changes in the telescope wavefront between observations. Therefore, the model of the residual starlight can be computed on local subsets of the image. The number of annular subsections can be specified with the \texttt{annuli} parameter, and likewise for radial subsections with the \texttt{subsections} parameter. The number of eigenimages a given model is composed of is controlled by the \texttt{numbasis} parameter, which we refer to as the number of ``KL modes". 

\begin{figure*}
\gridline{\fig{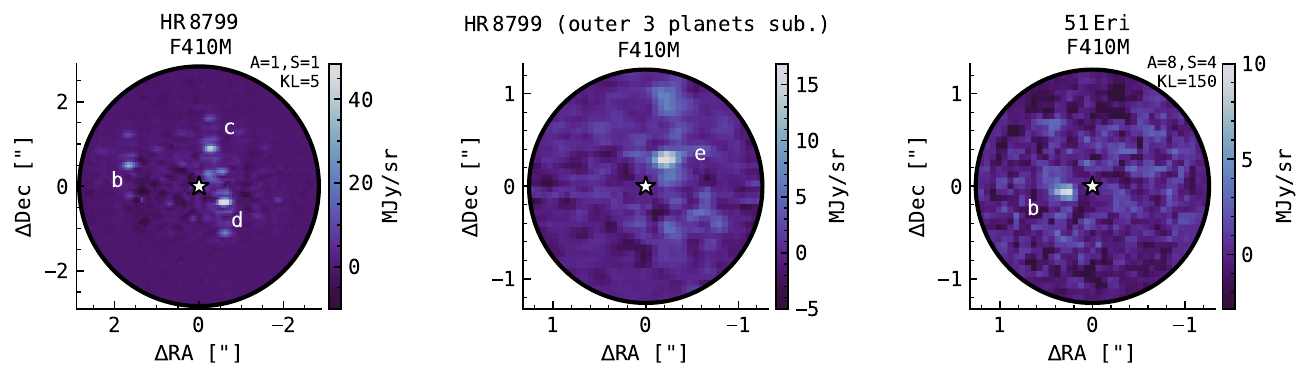}{\textwidth}{}}

\caption{Imaging Summary. \textit{Left:} PSF-subtracted F410M image of \hra, \texttt{pyKLIP} subtraction parameters labeled to the top right. \textit{Middle:} Same as left, but zoomed in to 1\farcs by 1\farcs, and the outer three planets have been forward modeled and subtracted, leaving \hre visible. See also the deconvolved image of \hra in this filter in Appendix \ref{appendix:deconv}, Figure \ref{fig:deconvolved}. \textit{Right:} PSF-subtracted F410M image of \cEb, showing the detection of the planet at its expected location.
    }\label{fig:imaging_summary}
\end{figure*}

\par In the longest-wavelength filters, namely F410M, F430M, and F460M, we detected all four \hra planets after subtracting the first principle component (that is, using KLIP parameters $\texttt{annuli}=1$, $\texttt{subsections}=1$, and $\texttt{numbasis}>1$). Including the angular diversity of the image set when constructing the model, that is using \texttt{ADI+RDI} mode, resulted in the self-subtraction of \hre due to its close separation and the strong dependence of the throughput at the location of the planet on the PA for each roll (the throughput degrades by up to 100\% for both \hre and \hrc from $\mathrm{PA}=92.33^\circ$ to $\mathrm{PA}=85.78^\circ$ depending on wavelength; see Figure \ref{fig:positions_mask}). We found that for our \hra observations, it was better to use \texttt{RDI} mode only and treat each roll separately to avoid self-subtraction from \texttt{ADI} and the large change in throughput on planets c,d,e between the two rolls (see Figure \ref{fig:positions_mask}) when extracting the forward modeled photometry (see below).

\par For our \cE observations, we are unable to recover the planet using the simplest set of KLIP parameters. There is a persistent residual speckle pattern that scales with wavelength across filters (which, anecdotally, appears to resemble a four-leaf clover). This could be due to a mismatch in centering between the target and reference (which is relatively small, ${\sim}5\,\mathrm{mas}$), or a difference in stellar magnitude or color. The difference in magnitude between HD~30562 and \cE is small, $\Delta K_{\rm mag}{=}0.2$, but their spectral types are noticeably distinct: G2IV versus F0V, which could create differences in the PSF, especially between 4 and 5\um where the G2-type reference star would have stronger CO absorption. We were able to recover \cEb in the F410M filter using a more aggressive set of KLIP parameters that split the image into four subsections and eight annular zones, using 50 eigenimages, but this subset of parameters still results in nondetections in the other filters. Qualitatively, this makes sense, as the first of eight annular zones restricts the starlight subtraction to the inner 0\farcs5, optimizing the subtraction where the planet is, and the four subsections split the subtraction above and below the bar, and to either side of the bar. This allows the variously sampled instances of the PSF that best represent the centering up and down the bar to contribute to the starlight subtraction at the location of the planet, and avoids trying to match the PSF on either end of the wedge with the symmetric features shared between both sides. Future work could conduct a more exhaustive search of KLIP parameter space \citep[e.g.,][]{AdamsRedai2023}, and including observations of multiple reference stars in this mode could potentially yield a better contrast performance, but in a coarse search, we found no indication that the contrast performance of this dataset could be improved to the point of robustly detecting \cEb in the other filters. Nevertheless, these nondetections set upper limits that inform the vertical mixing in the planet's atmosphere (see \S\ref{sec:analysis}). 

\par We used the Bayesian KLIP Astrometry (BKA) forward modeling functionality in \texttt{pyklip} to determine the planet's astrometry and photometry \citep{Wang2016}. We modeled the position-dependent PSFs of the four \hra planets in each roll angle and each filter independently using the \texttt{RDI} starlight subtraction. We subtracted the outermost best-fitting PSF from the data before continuing inward (so that, for instance, the forward model of \hre is fit to images containing only the PSF from \hre and the residuals from the fit to the outer three planets). We modeled the PSF of \cEb in the combined \texttt{ADI+RDI} sequence. We used the predicted positions of the planets as described in \S\ref{subsec:observing_strategy}. We generated \texttt{webbpsf} models of the off-axis coronagraphic PSF at our guess position. To account for the transmission of the planet PSF through the coronagraph, we compute this model twice, with and without the mask, and take the ratio of the integral under each PSF as the coronagraphic transmission for that position. We scale this model to have a contrast in the given filter of $5\times10^{-5}$ (see below for details regarding the contrast and flux determination); this model is then projected on the KLIP basis. We used the affine-invariant Markov Chain Monte Carlo (MCMC) algorithm from the \texttt{emcee} package \citep{Foreman-Mackey2013}, with 50 walkers taking 300 steps each, the first 100 of which were discarded as ``burn-in,'' to scale, translate, and subtract the model PSF from our images. In this process, we included a correlated noise term parameterized by a Gaussian process with a Mat\'ern $\nu=3/2$ kernel to account for residual speckle noise when estimating the astrometry and photometry \citep{Wang2016}. For the \hra planets, we averaged the two photometric measurements from each roll angle weighted by their uncertainties, and propagated uncertainties in quadrature. For the astrometry of these planets, we took the median and standard deviation of the astrometry measured across the filters. Our astrometry and photometry for each planet is recorded in Table \ref{tab:bka}. We show a summary of the PSF forward modeling (data, model, and residuals) in Appendix \ref{appendix:imaging}, Figure \ref{fig:bka}. 

\par Because there are no ``out-of-mask" stellar observations in each filter, in order to transform the photometric estimates from detector flux-calibrated units to units of contrast (to measure the planet's $\Delta\mathrm{mag}$, or to estimate the contrast performance of the mode in the next Section), we assume a stellar magnitude in the given filter by integrating under a synthetic stellar spectrum fit to archival photometry. We use \texttt{BT-NextGen} model spectra  \citep{Allard2011} for \hra~A and \cE. For \hra~A, we adopt parameters $T_{\rm eff}=7200\,\mathrm{K}$, $\log(g)=4.5$, and $\mathrm{[Fe/H]}=-0.5$, from \citet[][]{WangJi2020}, who found $7390\pm80$, $4.35\pm0.07$, and $-0.52\pm0.08$ from their joint analysis of LBT/PEPSI and HARPS high-resolution spectroscopy. For \cE~A, we adopt $T_{\rm eff}=7300\,\mathrm{K}$, $\log(g)=4.0$, and $\mathrm{[Fe/H]}=-0.1$, from \citet{Rajan2017}. We scaled these models to archival photometry from Gaia \citep{GaiaCollaboration2022}, Hipparcos/Tycho2 \citep{Hog2000}, and the Two Micron All Sky Survey \citep[2MASS;][]{Cutri2003, Skrutskie2006} using the \texttt{species} package \citep{Stolker2020}, minimizing the $\chi^2$ between the model and observations with the affine-invariant MCMC algorithm from the \texttt{emcee} package. Note that only our contrast, and not our apparent magnitude or flux density measurements, depend on this model assumption.

\par In summary, we achieved the program's goal of recovering all four \hra planets, notably the innermost planet \hre, in the four longest-wavelength filters F335M, F410M, F430M, and F460M (see Table \ref{tab:bka}, and Figure \ref{fig:imaging_summary}). This makes our images the first detection of \hre at 4.6\um, (our F460M being essentially equivalent to the \textit{Ms} filter used in observations from the ground; \citealt{Galicher2011}). We recover the outer three planets (\hrd, \hrc, and \hrb) in the F335M and F300M filters, and the outer two planets in all filters (including the LW channel F250M filter, and the F210M, F200W, and F182M filters projected onto the SW detector simultaneously during LW observations). We recover \cEb in the F410M filter, another first, but are unable to detect the planet in any other filter. We illustrate our detections across both systems in the F410M filter in Figure \ref{fig:imaging_summary}, and in Appendix \ref{appendix:imaging} Figures \ref{fig:img_summary_all_1} and \ref{fig:img_summary_all_2} for all other filters.

\begin{deluxetable*}{ccCCCCCCC}
\tablewidth{\textwidth}
\tablecaption{Relative astrometry and photometry in each filter for each planet measured in this work. \label{tab:bka}}
\tablehead{
\colhead{Planet} & \colhead{Filter} & \colhead{$\lambda_{\rm cen}$} & \colhead{$\mathrm{SNR}_{\rm bf, \mathcal{F}}$} & \colhead{$\Delta\mathrm{RA}$} & \colhead{$\Delta\mathrm{Dec}$} & \colhead{$\Delta\,\mathrm{mag}$} & \colhead{Apparent Magnitude} & \colhead{Flux density} \\
\colhead{} & \colhead{} & \colhead{[um]} & \colhead{} & \colhead{[ \farcs]} & \colhead{[ \farcs]} & \colhead{[$\mathrm{mag}$]} & \colhead{[$\mathrm{mag}$]} & \colhead{[\textmu Jy]} 
}
\startdata
\hrb & F182M & 1.845 & 5.2 & \cdots & \cdots & 13.50\pm0.11 & 18.66\pm0.11 &  28.8\pm3.0 \\ 
\hrb & F200W & 1.992 & 13.2 & \cdots & \cdots & 12.61\pm0.06 & 17.76\pm0.06 &  59.2\pm3.4 \\ 
\hrb & F210M & 2.097 & 14.6 & \cdots & \cdots & 12.24\pm0.06 & 17.39\pm0.05 &  75.4\pm4.1 \\ 
\hrb & F250M & 2.503 & 2.5 & \cdots & \cdots & 13.04\pm0.17 & 18.19\pm0.17 &  26.4\pm4.1 \\ 
\hrb & F300M & 2.993 & 6.4 & \cdots & \cdots & 12.22\pm0.04 & 17.37\pm0.04 &  41.4\pm1.6 \\ 
\hrb & F335M & 3.359 & 10.2 & \cdots & \cdots & 11.05\pm0.05 & 16.20\pm0.05 &  98.8\pm4.5 \\ 
\hrb & F410M & 4.084 & 18.0 & \cdots & \cdots & 10.10\pm0.04 & 15.25\pm0.04 & 164.6\pm6.7 \\ 
\hrb & F430M & 4.283 & 7.8 & \cdots & \cdots & 10.89\pm0.10 & 16.04\pm0.10 &  72.6\pm6.9 \\ 
\hrb & F460M & 4.631 & 6.6 & \cdots & \cdots & 10.86\pm0.14 & 16.01\pm0.13 &  64.5\pm8.3 \\ 
\hrb & all & \cdots & \cdots & 1.616\pm0.013 & 0.531\pm0.009 & \cdots & \cdots & \cdots \\
\hline 
% \hrc & F182M & 1.845 & 0.966 & \nodata        & \nodata       & \nodata      & \nodata      & \nodata      \\
\hrc & F200W & 1.992 & 4.1 & \cdots & \cdots & 12.31\pm0.13 & 17.46\pm0.13 &  78.1\pm9.7  \\
\hrc & F210M & 2.097 & 6.8 & \cdots & \cdots & 11.58\pm0.10 & 16.73\pm0.10 & 138.7\pm12.9 \\
\hrc & F250M & 2.503 & 4.5 & \cdots & \cdots & 11.80\pm0.20 & 16.95\pm0.20 &  83.1\pm26.5 \\
\hrc & F300M & 2.993 & 6.3 & \cdots & \cdots & 11.34\pm0.09 & 16.49\pm0.09 &  93.6\pm7.9  \\
\hrc & F335M & 3.359 & 12.3 & \cdots & \cdots & 10.19\pm0.12 & 15.34\pm0.12 & 216.8\pm23.5 \\
\hrc & F410M & 4.084 & 11.7 & \cdots & \cdots & 9.707\pm0.09 & 14.85\pm0.09 & 238.2\pm19.1 \\
\hrc & F430M & 4.283 & 9.2 & \cdots & \cdots & 10.04\pm0.05 & 15.19\pm0.05 & 159.7\pm7.7  \\
\hrc & F460M & 4.631 & 8.2 & \cdots & \cdots & 10.02\pm0.08 & 15.17\pm0.08 & 139.4\pm10.4 \\
\hrc & all & \cdots & \cdots & -0.291\pm0.012 & 0.911\pm0.003 & \cdots & \cdots & \cdots \\
\hline 
% \hrd & F182M & 1.845 & 1.726 & \nodata        & \nodata        & \nodata      & \nodata      & \nodata      \\
% \hrd & F200W & 1.992 & 2.481 & \nodata        & \nodata        & \nodata      & \nodata      & \nodata      \\
% \hrd & F210M & 2.097 & 2.827 & \nodata        & \nodata        & \nodata      & \nodata      & \nodata      \\
\hrd & F250M & 2.503 & 5.7 & \cdots & \cdots & 11.50\pm0.14 & 16.65\pm0.14 & 109.5\pm14.4 \\
\hrd & F300M & 2.993 & 12.5 & \cdots & \cdots & 10.63\pm0.09 & 15.78\pm0.09 & 179.4\pm16.2 \\
\hrd & F335M & 3.359 & 23.3 & \cdots & \cdots &  9.66\pm0.05 & 14.81\pm0.05 & 355.7\pm18.2 \\
\hrd & F410M & 4.084 & 31.0 & \cdots & \cdots &  9.11\pm0.04 & 14.26\pm0.04 & 410.5\pm16.7 \\
\hrd & F430M & 4.283 & 30.2 & \cdots & \cdots &  9.17\pm0.05 & 14.32\pm0.05 & 354.4\pm17.7 \\
\hrd & F460M & 4.631 & 25.8 & \cdots & \cdots &  9.12\pm0.06 & 14.27\pm0.06 & 319.2\pm18.8 \\
\hrd & all & \cdots & \cdots & -0.61\pm0.007 & -0.348\pm0.005 & \cdots & \cdots & \cdots \\
\hline 
% \hre & F182M & 1.845 & 1.465 & \nodata        & \nodata       & \nodata      & \nodata      & \nodata      \\ 
% \hre & F200W & 1.992 & 0.694 & \nodata        & \nodata       & \nodata      & \nodata      & \nodata      \\ 
% \hre & F210M & 2.097 & 0.131 & \nodata        & \nodata       & \nodata      & \nodata      & \nodata      \\ 
% \hre & F250M & 2.503 & 3.285 & \nodata        & \nodata       & \nodata      & \nodata      & \nodata      \\ 
% \hre & F300M & 2.993 & 1.464 & \nodata        & \nodata       & \nodata      & \nodata      & \nodata      \\ 
\hre & F335M & 3.359 & 5.8 & \cdots & \cdots & 10.34\pm0.16 & 15.49\pm0.16 & 188.8\pm28.0 \\ 
\hre & F410M & 4.084 & 9.95 & \cdots & \cdots &  9.62\pm0.16 & 14.77\pm0.16 & 256.6\pm38.6 \\ 
\hre & F430M & 4.283 & 6.1 & \cdots & \cdots & 10.06\pm0.12 & 15.21\pm0.12 & 156.3\pm17.8 \\ 
\hre & F460M & 4.631 & 5.9 & \cdots & \cdots &  9.87\pm0.11 & 15.03\pm0.11 & 159.7\pm17.6 \\
\hre & all & \cdots & \cdots & -0.226\pm0.014 & 0.332\pm0.015 & \cdots & \cdots & \cdots \\
\hline 
% \cEb & F335M & 3.359 & 0.8 & 0.266\pm 0.058 & -0.095\pm0.047 & 13.7\pm1.1 & 18.2\pm1.1 & 16.3\pm16.4 \\ 
\cEb & F410M & 4.084 & 4.7 & 0.286\pm0.010 & -0.099\pm0.004 & 11.4\pm0.1 & 15.8\pm0.1 & 95.9\pm9.8 \\ 
% \cEb & F430M & 4.283 & 3.4 & 0.283\pm0.013 & -0.094\pm0.009 & 11.8\pm0.3 & 16.3\pm0.3 & 59.7\pm14.7 \\ 
% \cEb & F460M & 4.631 & 2.2 & 0.261\pm0.027 & -0.084\pm0.019 & 11.9\pm0.5 & 16.4\pm0.5 & 46.7\pm20.6
\enddata
\tablecomments{$\mathrm{SNR}_{\rm bf, \mathcal{F}}$ as defined in \citet{Golomb2021} is the standard deviation of the nearby pixels compared to the peak flux on the planet; it is not calibrated to account for, e.g. small sample statistics or the algorithmic throughput, as is done for the contrast curve in Figure \ref{fig:f410m_contrast_curve_hr8799}. $\Delta\mathrm{RA}$, $\Delta\mathrm{Dec}$ are the position measurements relative to the central star, and $\Delta\mathrm{mag}$ is the difference in magnitude in a given filter for a planet relative to the synthesized magnitude for the central star. Apparent magnitude and flux density are measured as described in \S\ref{subsec:psfsub}, and given in vegamags and microjansky, respectively.}
\end{deluxetable*}

\subsection{Contrast estimation}

\par Accurate contrast estimation from our datasets is especially challenging, because the presence of close-in planets and their diffuse NIRCam coronagraphic PSFs mean that there are effectively very few pixels that capture pure instances of stellar speckle ``noise" inward of 1\farcs0 for the \hra observations, and 0\farcs5 for the \cE observations. Typically, pixels influenced by the planet's PSF would be subtracted, or masked, or both, before calculating some standard deviation of pixels within an annulus. Detecting the innermost planets at face value implies a certain degree of performance ($C{=}1\times10^{-4}$ at 0\farcs4, based on our ${\sim}10\,\sigma$ detection of \hre in the F410M filter), but more challenging is quantifying the general contrast performance of the mode, and comparing it ``apples-to-apples" with other modes and instruments. Similarly challenging to consider is our apparent ``detection" of \cEb, given the appearance of a point-source at the anticipated location and brightness of the planet, but that falls beneath our $5\,\sigma$ calibrated contrast curve ($4.7\,\sigma$, comparing peak pixel count to the standard deviation of noise within a 20 pixel stamp, but $<3\,\sigma$ accounting for small sample statistics following the correction in \citealt{Mawet2014}). This small sample statistics correction is applied when the null hypothesis is that there \textit{does not} exist a planet within a given annulus, which is absolutely necessary to reject false positives in the case of an unbiased planet search from the ground for instance, but is not necessarily applicable to the case of observing a known planet from a relatively much more stable space telescope. Although we use the constraint provided by the \cE observations in our analysis, without the prior predictive orbits and atmospheres that indicate the point-source in Figure \ref{fig:imaging_summary} is \cEb, it would be unwise to claim a detection of a new planet based solely on this kind of dataset. This further highlights the strengths and weaknesses of the mode---it is a very effective characterization tool, but with limited discovery space. The reliability of detections at the ``bleeding edge" of the wedge can be bolstered by a priori information, as we argue is the case for our detection of \cEb (for instance, strong position predictions from previous relative astrometry or from joint radial velocity and proper-motion anomaly orbit fits; see the example of Eps Indi b detected with JWST/MIRI, \citealt{Matthews2024}). Nevertheless, much of the interesting science to be done with this mode is at or below the small sample statistics corrected contrast threshold. Discoveries of newly imaged planets using this mode will require careful interpretation and diligent follow-up. 
\par
\begin{figure}
    \centering
    % \gridline{

    % \gridline{\fig{figures/nircam_f410m_planetsubbed_5sigma_cc_logx.pdf
    % }{\textwidth}{Contrast curves for the F410M filter}}

    % \gridline{\fig{figures/HR8799_LWBAR_all_lw_filter_cc.pdf}{\textwidth}{Contrast curves for the entire set of \hra observations, across all filters}}
    
    \includegraphics[width=\linewidth]{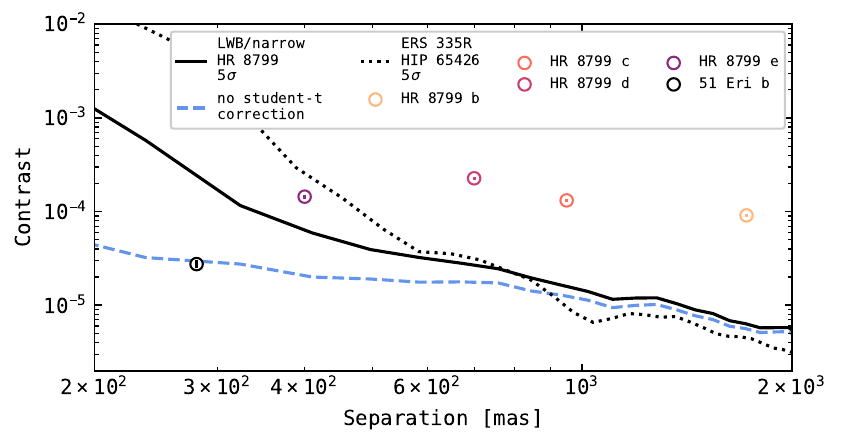}
    
    \caption{JWST/NIRCam F410M $5\,\sigma$ contrast curves for the \texttt{LWB}/narrow coronagraph (solid black line) and the \texttt{335R} coronagraph from \citep[dashed black line;][]{Carter2023} The turn over in contrast performance between the two modes occurs at about 0\farcs75 (just outside of the IWA for the \texttt{MASK335R}). Our detections of the 4 \hra planets, recorded in Table \ref{tab:bka}, are shown with their associated contrast uncertainty as an error bar inset within their scatter points. The wedge mask outperforms the round mask at close separations (${<}$\,0\farcs7) thanks to its sharper inner working angle and higher throughput, but suffers at longer separations. The \texttt{LWB}/narrow mode achieves $C_{4.1\um}=10^{-4}$ at 0\farcs33 at $5\,\sigma$.}
    \label{fig:f410m_contrast_curve_hr8799}
\end{figure}

 We adapt the contrast estimation tools in \texttt{spaceKLIP} for the NIRCam \texttt{LWB} in the following ways. The best-fit \texttt{webbpsf} forward models of the four planets following the procedure in \S\ref{subsec:psfsub} were subtracted from the pre-starlight-subtracted dataset. Then, we performed starlight subtraction as usual and masked $1.5{\times}\mathrm{FWHM}$ pixels centered on the location of each planet. We also masked and replaced the wedges of the image where the attenuation of the bar is greatest, considering only pixels perpendicular to the mask and not, for instance, underneath it \citep[see Figure 6 in][]{Kammerer2022}. Then, following standard \texttt{pyKLIP}/\texttt{spaceKLIP} procedure, we estimate the annular standard deviation in the remaining pixels, accounting for small sample statistics using the Student's t-distribution correction \citep{Mawet2014}. The result for the F410M filter is visualized in Figure \ref{fig:f410m_contrast_curve_hr8799}; the planets are overlaid, in addition to the equivalent contrast curve from the ERS program \citep{Carter2023}, using the same filter but observed using the round \texttt{MASK335R} coronagraph. Appendix \ref{appendix:imaging}, Figure \ref{fig:all_ccs} shows the contrast for each filter and detection across the \hra observations.

\section{Analysis}
\label{sec:analysis}

\subsection{\hra} \label{subsec:hr8799_analysis}

\par As a multiplanet system following the low-gravity, late L-type sequence, the \hra planets are necessarily difficult to model succinctly. Orbital modeling needs to account for five bodies and their potential interactions while relying on fractional orbit coverage. The spectrum of each planet is shaped by thick clouds and non-equilibrium chemistry, and requires understanding the calibration of multiple instruments across two decades of observation.
\par Given that an orbital analysis of the high-precision VLTI/GRAVITY observations of the HR~8799 system is forthcoming (A. Chavez, priv. comm.), we do not update the orbital solution for the system based on our new relative astrometry, which has much lower precision compared to ground-based observations at shorter wavelengths. We refer the reader to \citet{Zurlo2022, Thompson2023} for two recent examples of the orbital solution for the system.
\par
\begin{figure*}
    \centering
    \gridline{\fig{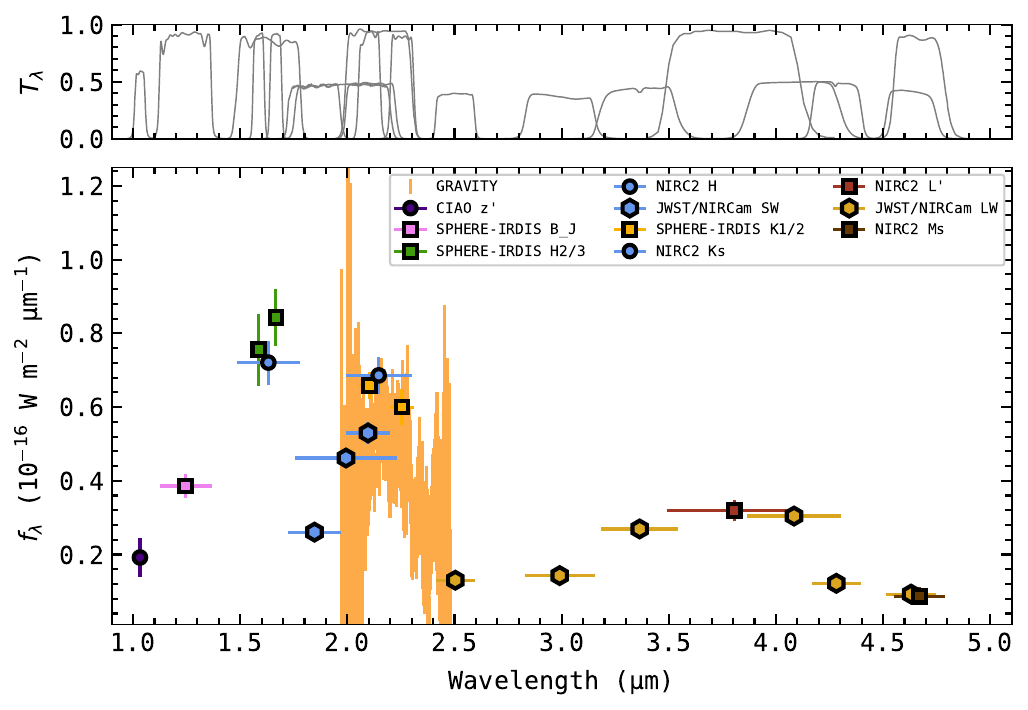
    % \gridline{\fig{figures/retrieval_comparison/HR8799b_nasedkin24_compare_alldata.pdf
    }{0.48\textwidth}{\hrb} 
    \fig{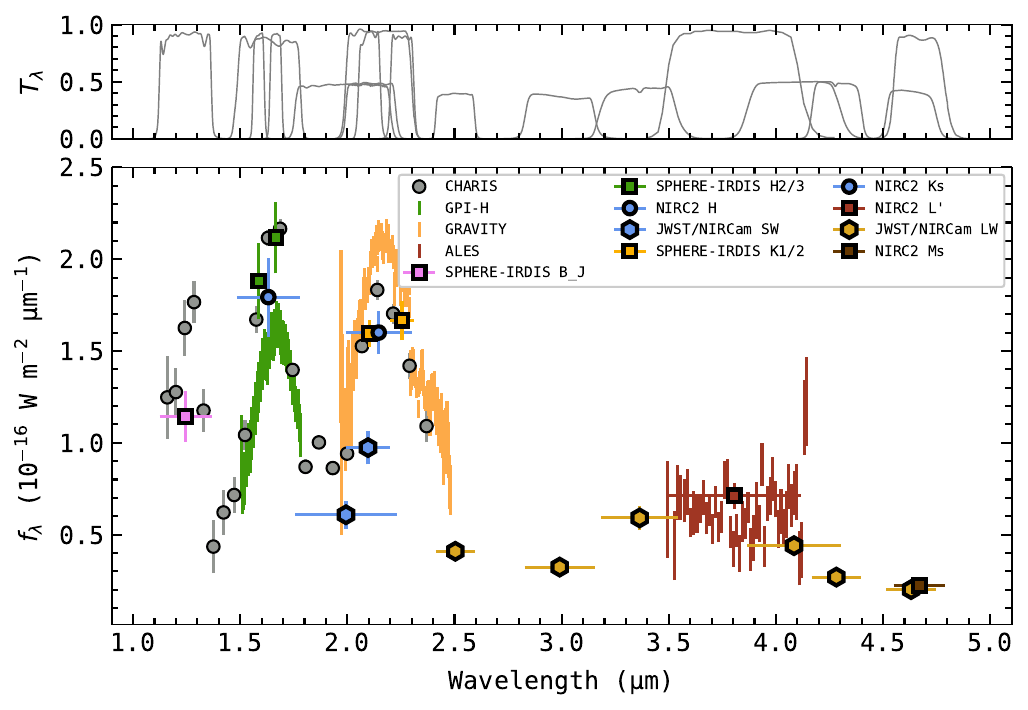
    % \fig{figures/retrieval_comparison/HR8799c_nasedkin24_compare_alldata.pdf
    }{0.48\textwidth}{\hrc} 
    }
    \gridline{
    \fig{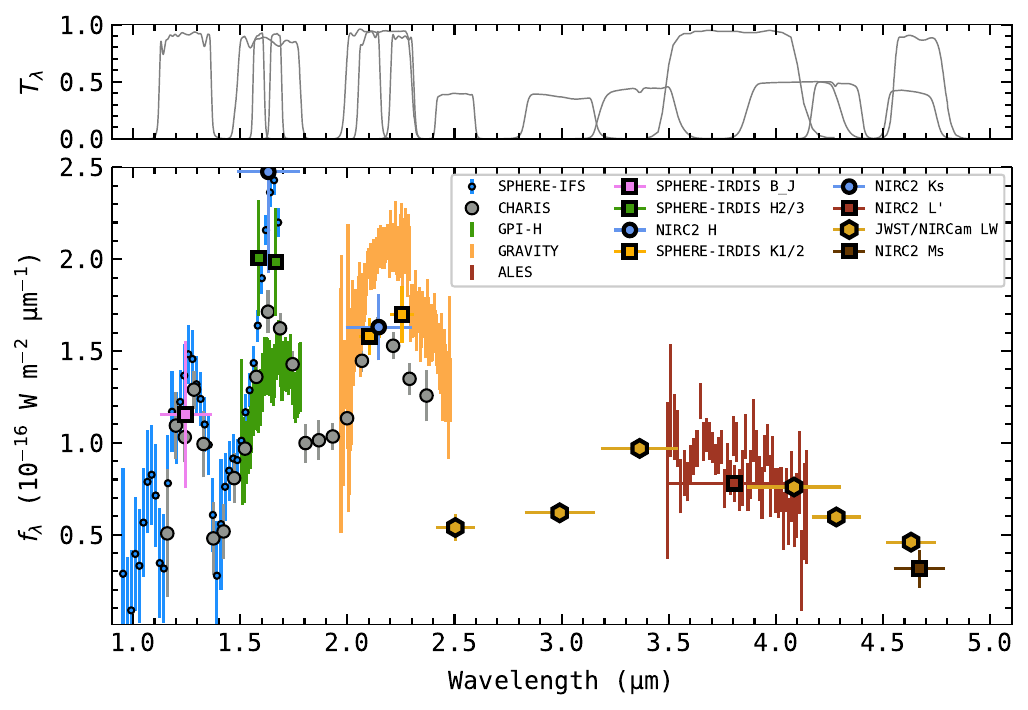
    % \fig{figures/retrieval_comparison/HR8799d_nasedkin24_compare_alldata.pdf
    }{0.48\textwidth}{\hrd}
    \fig{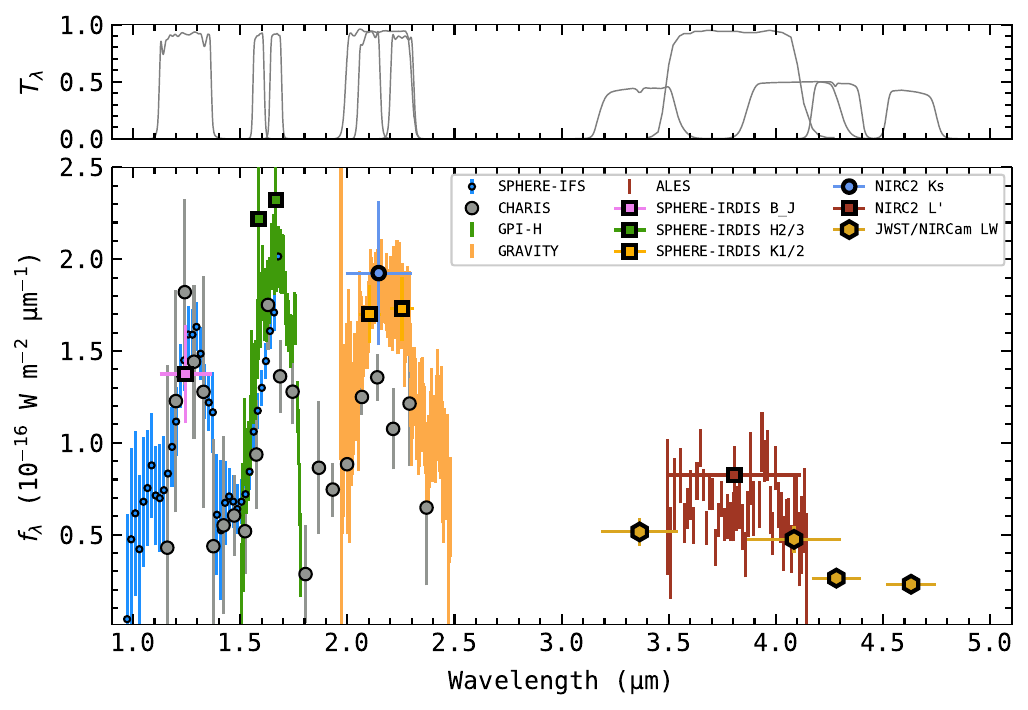
    % \fig{figures/retrieval_comparison/HR8799e_nasedkin24_compare_alldata.pdf
    }{0.48\textwidth}{\hre}
    }
    
    \caption{\texttt{LWB}/narrow observations of all four \hra planets (SW filters as blue hexagons, LW filters as gold hexagons) compared to the ground-based observations of the planets collated by \citet{Nasedkin2024} and used in their retrieval analysis \citep[data from][]{Marois2008, Fukagawa2009, Bergfors2011, Galicher2011, Esposito2013, Konopacky2013, Currie2014, Skemer2014, Barman2015,  Maire2015, Greenbaum2018, GRAVITYCollaboration2019, Doelman2022, Wang2022, Nasedkin2024}. A comparison with their models can be found in Appendix \ref{appendix:retrieval_comparison}, Figure \ref{fig:all_compare_retrievals}. Our measurements show agreement with groundbased \textit{L}- and \textit{M} band observations to within $1-2\,\sigma$. There is tension in the \textit{K} band between our observations at F182M, F200W, F210M, and F250M and the VLTI/GRAVITY spectra, especially of \hrc and \hrd. This could be due to the absolute flux calibration of the VLTI/GRAVITY spectra, see discussion in \S\ref{subsec:hr8799_analysis}. The variation in 3-4\um and 4-5\um slopes between each planet indicate unique photospheric abundances of CH$_4$, CO, and CO$_2$, as discussed in \S\ref{subsec:hr8799_discussion}}
    \label{fig:compare_datasets}
\end{figure*}

 Then, given the recent comprehensive atmospheric modeling analysis presented in \citet{Nasedkin2024} and the complexity (and computational expense) of the atmospheric modeling involved to satisfactorily fit the spectral energy distributions (SEDs) of the \hra planets, we defer the atmospheric modeling of our new \hra observations to future work. Figure \ref{fig:compare_datasets} compares the variety of ground-based observations collated by \citep{Nasedkin2024} and used in their \texttt{petitRADTRANS} atmospheric retrieval analysis; in Appendix \ref{appendix:retrieval_comparison}, Figure \ref{fig:all_compare_retrievals} compares the retrievals from \citet{Nasedkin2024} fit to the ground-based data with synthesized and observed photometry in our NIRCam filters. 

Offsets are apparent between many of the retrieved models and our data, particularly between 2 and 3\um. This could be due to the decision in \citet{Nasedkin2024} to fix the absolute flux of the VLTI/GRAVITY spectra and scale most of the other ground-based spectra to match the best-fitting models using offset parameters. The VLTI/GRAVITY spectra may have some systematic offset. This is possible, since unlike instruments with dedicated satellite spots or other methods for continual flux calibration, some exoplanet observing strategies using GRAVITY, like the dual field off-axis mode \citep{Nowak2024} that was used to observe the outer \hra planets, do not contemporaneously monitor the flux of the host throughout the night, instead taking absolute flux calibration observations at the start or end of a sequence. This is because these observations require metrology calibration from a binary star after observing the planet in order to determine the zero-point of some interferometric observables \citep[see \S2.1 in][]{Nowak2024}, whereas the absolute flux requires swapping the observing mode and therefore scrambling the metrology information. Future observations can leverage the improved adpative optics performance of the GRAVITY+ upgrade to observe the inner \hra planets in ``on-axis" mode, where more contemporaneous data on the host star (and therefore more accurate flux calibration) can be obtained. Nevertheless, our qualitative results generally agree with the findings in \citet{Nasedkin2024}, which we discuss below.

\par If there was a significant flux calibration issue with the JWST photometry, we'd expect to see significant offsets between equivalent filters. This could manifest in two ways, either (1) a systematic offset across all filters or (2) offsets/disagreements varying by planet or filter, perhaps depending on the coronagraphic throughput correction. We can rule out the first case by noting the excellent agreement between \hrb (with a wide enough separation to have 100\% coronagraphic and algorithmic throughput) and literature photometry. Especially notable is the excellent agreement between the three outer planets and their \textit{M} band photometry at 4.6\um \citep{Hinz2010, Galicher2011}; because all of the models in \citet{Nasedkin2024} were fit to this photometric point, which provides unique information about the CO abundance as a function of pressure, all the retrievals thread through this point (and therefore agree with our F460M photometry). Regarding the second possibility, our NIRCam photometry agrees well within uncertainties with LBT/ALES low-resolution spectroscopy \citep{Doelman2022}, as well as the \textit{L'} band photometry \citep{Marois2008, Marois2010, Hinz2010, Currie2011, Currie2014, Thompson2023}. There is more significant tension between the assorted narrowband LBT/LMIRCam photometry \citep{Skemer2014}, but this appears driven by the large variation and lower significance detections in those data, as there is not a systematic offset for all planets (for instance, some of the 3\um narrowband photometry of \hrd are significantly lower flux than the NIRCam photometry or ALES spectra, whereas the \hrc fluxes are significantly higher). The narrowband, ground-based photometry of the planets at Br-$\alpha$ from \citet{Currie2014} disagrees with and is overluminous by $1-2\,\sigma$ compared to our overlapping, but higher precision F410M photometry. Looking forward, our 1.8-5\um space-based photometry of this system can help anchor the absolute flux scaling of spectroscopic observations for atmospheric analyses.

% all four planets

% single planet figure
% \begin{figure*}
%     \centering
%     \includegraphics[width=\linewidth]{figures/HR8799c_nasedkin24_compare.pdf}
%     \caption{\texttt{LWBAR}/narrow observations of \hrc (SW filters as blue hexagons, LW filters as gold hexagons) compared to \texttt{petitRADTRANS} models fit to ground-based data from \citet{Nasedkin2024}. Model ``c.AB.9" with one of the bluest F410M-F430M colors is shown in black, and synthesized photometry from this model for each NIRCam filter considered is shown as an open hexagon; all other models from the exhaustive retrieval analysis are plotted in purple. Some of the retrieval models are consistent with some of the NIRCam photometry, but many are not, especially at shorter wavelengths. This could be due to the absolute flux calibration of the ground-based spectra, some of which were scaled with varying offset parameters.}
%     \label{fig:c_compare_retrievals}
% \end{figure*}

\par
\begin{figure*}
    \centering
    \includegraphics[width=0.9\linewidth]{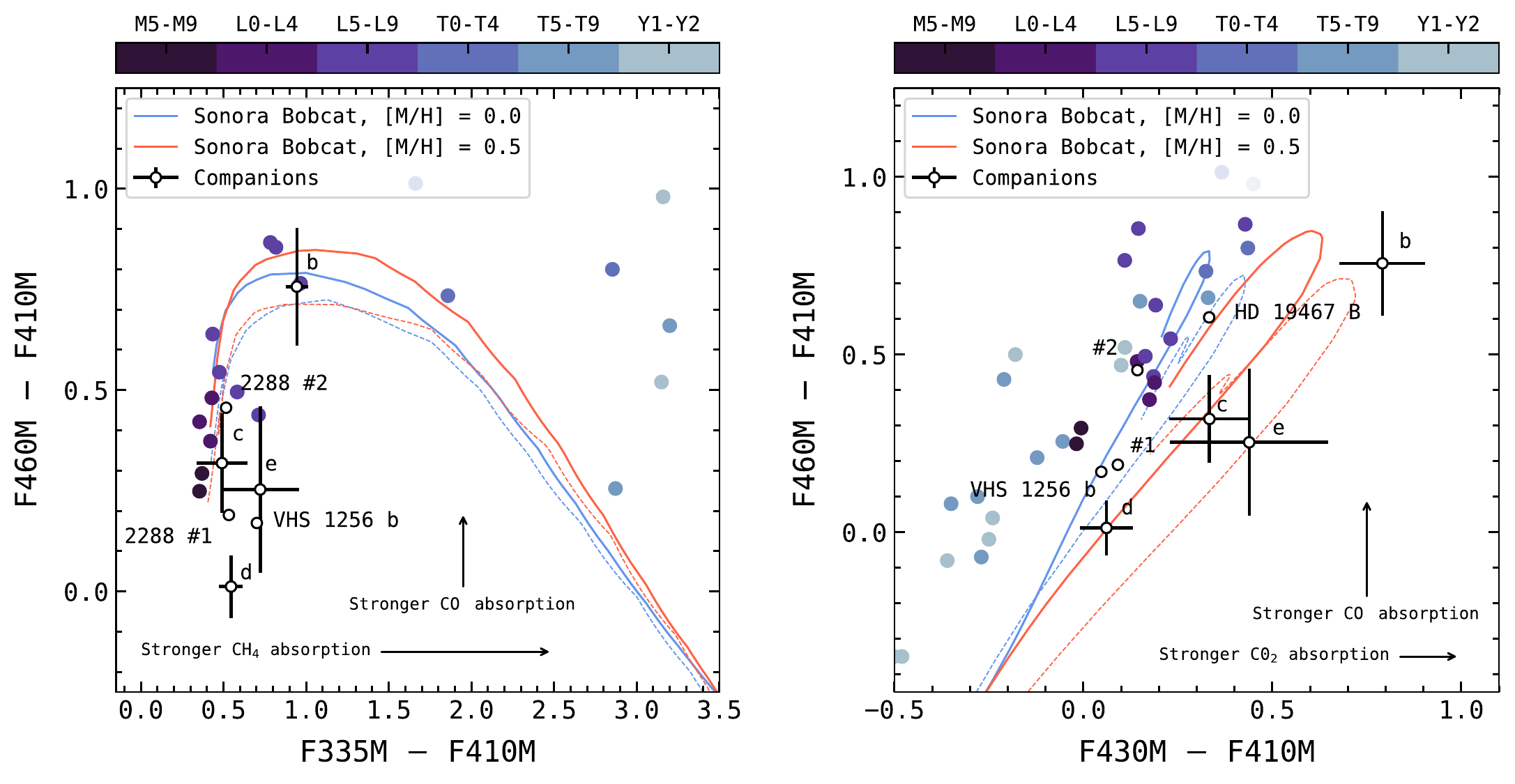}
    \caption{Color-color diagrams with NIRCam 3-5\um medium-band filters. Companions, namely the \hra planets, the field-age BD companion HD~19467~B \citep{Greenbaum2023, Hoch2024}, and the young companion VHS~1256~b \citep{Miles2023} are illustrated in black. Also shown in black are the two field L-dwarfs targeted by GO 2288 (PI: J. Lothringer); the BD labeled `\#1' is 2MASSW J2148162+400359, and `\#2' is 2MASS J06244595-4521548. The AKARI sample of L-T field BDs \citep{Sorahana2013} and the JWST sample of late T- and Y-dwarfs \citep{Beiler2024} are colored by their spectral type. Over plotted are cloudless \texttt{Sonora-Bobcat} evolutionary models \citep{Marley2021} at 1~Gyr (solid) and 30~Myr (dashed) for solar metallicity (blue) and enhanced metallicity (orange). A number of qualitative features are noticeable, (1) the \hra planets are offset towards bluer F430M-F410M colors than the field sample, corresponding to the younger and more metal enhanced evolutionary model tracks, (2) the planets show significant spread across the diagram despite their apparently similar photometric colors in the NIR \citep[see, e.g., Figure 1 in ][]{Bonnefoy2016}, and (3) there is an apparent sequence, from the inner planet \hrd to the outer planet \hrb in F460M-F410M color.}
    \label{fig:color-color}
\end{figure*}

 In this work, our primary goal is making these new photometric observations available and conducting a qualitative analysis based on the color of the \hra planets, as opposed to exhaustively modeling their atmospheres. We compared the 3-5\um colors of the \hra planets to the sample of field L- and T-type BDs observed by the AKARI spacecraft \citep{Sorahana2012, Sorahana2013} and the sample of field T- and Y-type BDs observed by JWST \citep{Beiler2024}. Figure \ref{fig:color-color} shows a F430M-F410M versus F460M-F410M color-color diagram for the field sample and the \hra planets, with cloudless \texttt{Sonora-Bobcat} evolutionary models. The qualitative comparison with a cloudless model is acceptable despite the strong cloud opacity in the planets' atmospheres because at these wavelengths the clouds are effectively ``gray," meaning wavelength independent. These color-color diagrams reveal three interesting qualitative results that we discuss in \S\ref{sec:discussion}. First, the \hra planets are offset toward bluer F430M-F410M colors than the field BDs by about half a magnitude. Second, despite their relatively similar NIR colors, the \hra planets are dramatically spread (by up to a magnitude) in their 3-5\um colors. Third, the \hra planets follow an apparent radial sequence in F460M-F410M color (which traces CO absorption), with planet d being the reddest close in, and planets c and b becoming increasingly blue.

\subsection{\cE} \label{subsec:51eri_analysis}

\par The orbit and atmosphere of \cEb are much more tractable to model within the scope of this paper. The system architecture necessitates only a two-body Keplerian orbit fit, and the effective temperature range of the planet means that its observed spectrum is not as strongly influenced by cloud opacity as the \hra planets, meaning that ``out of the box" precomputed grids of atmospheric models will be sufficient to capture the spectral variations in the observations.

\subsubsection{Orbit}

\begin{figure}
    \centering
    \includegraphics[width=\linewidth]{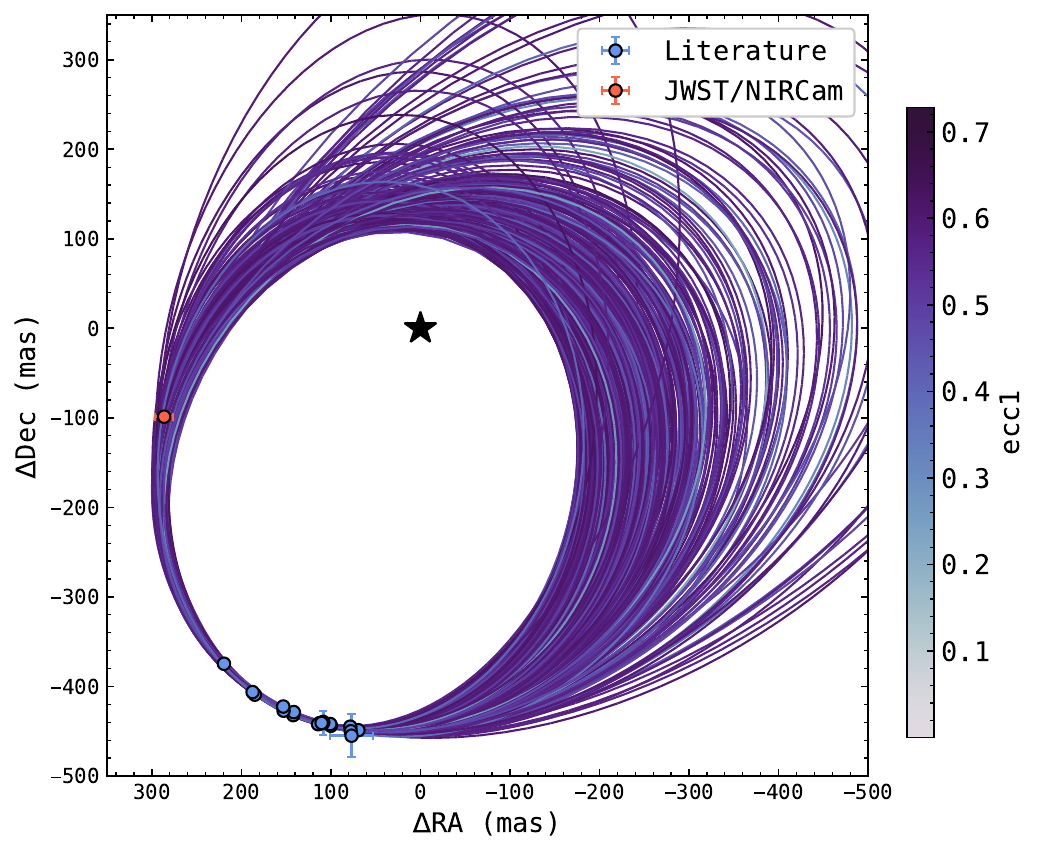}
    \caption{The orbit of \cEb. Five-hundred randomly drawn orbits from our posterior distribution, determined using \texttt{orbitize!} in \S\ref{subsec:51eri_analysis}. Scatter points in blue represent literature relative astrometry, and those in red represent our new JWST/NIRCam measurement. The new detection agrees well with high-eccentricity ($e=0.58$) solutions fit to previous observations.}
    \label{fig:51eri_orbit}
\end{figure}

\par Our detection of 51~Eri~b is the newest published sighting of the planet since late 2018, and as expected, the planet appears significantly further along in its orbit. We updated the orbital solution for the planet using the python package \texttt{orbitize!}\footnote{\url{https://orbitize.readthedocs.io/en/latest/}} \citep{Blunt2020}. We compiled the relative astrometry of the planet from Gemini/GPI and Keck/NIRC2 \citep{Macintosh2015, DeRosa2015, DeRosa2020}, VLT/SPHERE \citep{Maire2019}, and our new JWST/NIRCam measurement, as well as the absolute astrometric measurements of the host star from Hipparcos and Gaia in the eDR3 version of the Hipparcos-Gaia Catalogue of Accelerations \citep[HGCA,][]{Brandt2021}. We set a normally distributed prior on the system parallax equivalent to the Gaia DR3 measurement, $\pi\in\mathcal{N}(\mu=33.439, \sigma=0.077)\,\mathrm{mas}$ \citep{GaiaCollaboration2022}. We also set a normally distributed prior on the primary star's mass $M_\star\in\mathcal{N}(\mu=1.550, \sigma=0.006)\,M_\odot$, based on the updated stellar radius measurement and model analysis in \citet{Elliott2024}. We sampled orbits using the parallel tempered, affine-invariant \texttt{ptemcee} MCMC algorithm \citep{Foreman-Mackey2013, Vousden2016}. We used MCMC chains with 20 temperatures and 1000 walkers, which were run for 20,000 steps in an initial ``burn-in" period. Then, each walker was run for 20,000 steps, where every 10th step was recorded. The lowest-temperature chains for each walker comprise our final posterior estimate, with 2,000 by 1,000 = 2,000,000 accepted orbits. Table \ref{tab:orbit} records the posterior distribution on orbital parameters from this orbit fit. Figure \ref{fig:51eri_orbit} shows 500 orbits randomly drawn from the posterior distribution, which is visualized as a corner plot in Appendix \ref{appendix:corner_plots} Figure \ref{fig:51eri_orbit_corner}. Our JWST/NIRCam detection confirms the preference for high-eccentricity orbits that was found previously \citep{Maire2019, DeRosa2020, Dupuy2022}. We discuss the updated orbit in more detail in \S\ref{sec:discussion}.

\begin{deluxetable}{ccC}
\tablewidth{\textwidth}
\tablecaption{Orbital parameters inferred for \cEb in this work. \label{tab:orbit}}
\tablehead{
\colhead{Parameter} & \colhead{Prior} & \colhead{Posterior}
}
\startdata
$a$ [au]                              & $\log\,\mathcal{U}(0.001, 1e4)$   & 9.58_{-0.42}^{1.61} \\
$e$                                   & $\mathcal{U}(0,1)$                & 0.57_{-0.09}^{0.03} \\
$i$ [$^\circ$]                        & $\mathrm{Sine}(0,\pi)$            & 151.1_{-11.8}^{5.3} \\
$\omega$ [$^\circ$]                 & $\mathcal{U}(0,2\pi)$             & 96.8_{-38.1}^{{12.7}^\dag} \\
$\Omega$ [$^\circ$]                 & $\mathcal{U}(0,2\pi)$             & 94.9_{-51.0}^{{26.6}^\dag} \\
$\tau^\ddag$                        & $\mathcal{U}(0,1)$                & 0.30_{-0.06}^{0.05} \\
$\pi$ [mas]                         & $\mathcal{N}(33.439, 0.077)$      & 33.43_{-0.08}^{0.08} \\
$M_\mathrm{b}$ [$M_\mathrm{J}$]     & $\log\,\mathcal{U}(0.1,100)$      & 0.88_{-0.67}^{{+2.71}^\ddag} \\
$M_\mathrm{A}$ [$M_\odot$]          & $\mathcal{N}(1.550,\,0.006)$      & 1.55_{-0.01}^{+0.01}
\enddata
\tablecomments{We report the median and 68\% confidence interval on each parameter derived from the posterior solution of orbits. $^\dag$ A degenerate solution, with equally likely peak modulo $180^\circ$. $^\ddag$ $\tau$ is a dimensionless quantity describing the periastron passage of the orbit, $\tau=(t_{\rm peri}-t_{\rm ref})/P$, where $t_{\rm ref}=58849\,\mathrm{MJD}$. $^\ddag$This is effectively an upper limit on mass, with $M_{\rm b}{<}9.0\,M_\mathrm{J}$ at $3\,\sigma$}
\end{deluxetable}

\subsubsection{Atmosphere}

\par We compared the observed SED of \cEb to the cloudy, radiative convective equilibrium (RCE) atmosphere model grid \texttt{Exo-REM} \citep{Charnay2018}. This model was chosen because it includes the effects of vertical mixing induced disequilibrium chemistry and determines the particle size distribution of the iron and silicate clouds using a simplified microphysical model. The available model grid\footnote{\url{https://lesia.obspm.fr/exorem/YGP_grids/}} varies atmospheric temperature, surface gravity, and abundances (parameterized by metallicity, [M/H], and the carbon-to-oxygen ratio, C/O). We used the python package \texttt{species}\footnote{\url{https://species.readthedocs.io/en/latest/}} \citep{Stolker2020} to handle the model comparison; \texttt{species} provides a user-friendly interface for grid interpolation and model fitting. \texttt{pyMultinest} \citep{Buchner2014}, a python interface for \texttt{Multinest} \citep{Feroz2008, Feroz2009, Feroz2019}, was used to estimate the posterior distribution of the parameters of interest ($T_\mathrm{eff}$, $\log(g)$, $R$, [Fe/H], C/O) with nested sampling.

\begin{figure*}
    \centering
    \includegraphics[width=\textwidth]{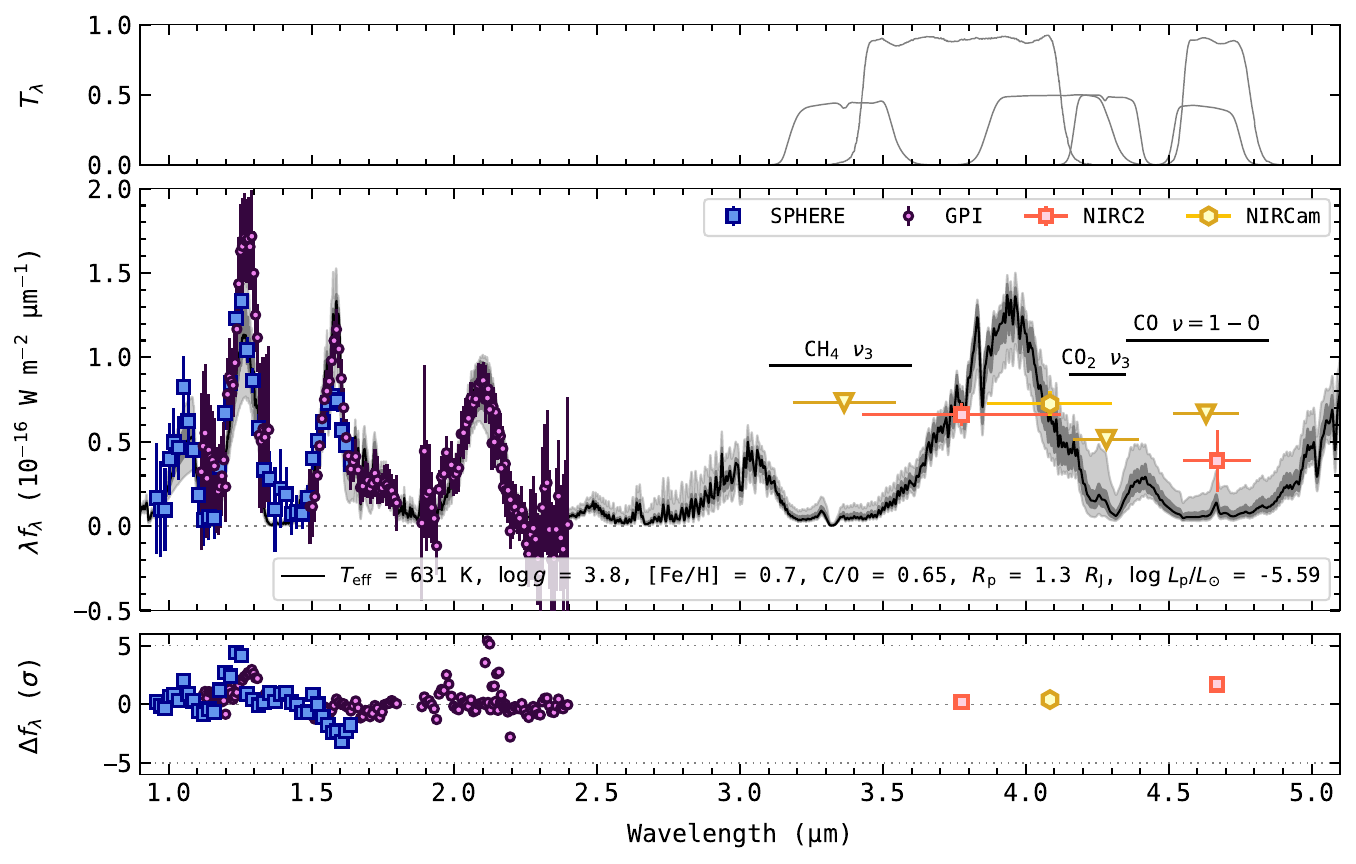}
    \caption{SED of \cEb and best-fitting \texttt{Exo-REM} models. The top panel illustrates the transmission functions for each photometric filter. The middle panel shows the GPI \citep[black circles,][]{Macintosh2015, Rajan2017} and SPHERE \citep[black squares,][]{Samland2017, Brown-Sevilla2023} low-resolution spectra of the planet, as well as ground-based \textit{L'}  and \textit{M} band photometry \citep[blue squares,][]{Rajan2017}, and the measurements from this work (gold hexagon, for our F410M detection, and gold downward facing triangles for our F335M, F430M, and F460M nondetections). Over plotted is the best-fitting \texttt{Exo-REM} model in magenta, and then $1,2\,\sigma$ envelopes showing the range of accepted models. The bottom panel illustrates the residuals to the best-fitting model; there are noticeable residuals in the \textit{J} band GPI spectrum and the \textit{H} band SPHERE spectrum that could be related to speckle noise or instrument throughput. The fit shows good agreement to the \textit{L'} band and F410M photometry, which constrain the CH$_4$ and CO$_2$ absorption slopes. The \textit{M} band photometry from \citet{Rajan2017} lies slightly above the best-fitting model, but in general is consistent with strong CO absorption due to vertical mixing. }
    \label{fig:51erib_exorem_spectrum}
\end{figure*}

\par Our dataset included the \textit{J}-, \textit{H}-, and \textit{K1} and \textit{K2}-band integral field unit (IFU) spectra of the planet from Gemini/GPI \citep{Rajan2017}, the \textit{Y/J/H1}-band IFU spectra from VLT/SPHERE \citep{Brown-Sevilla2023}, the \textit{L'}  and \textit{M} band photometry from Keck/NIRC2 \citep{Rajan2017}, and our new JWST/NIRCam F410M photometric point. We set a normally distributed prior on the system parallax equivalent to the Gaia DR3 measurement, $\pi\in\mathcal{N}(\mu=33.439, \sigma=0.077)\,\mathrm{mas}$; the sampler draws from $R$ and $\pi$ to scale the absolute flux of the model to match the observations. The covariance of the low-resolution spectra are estimated using Gaussian Process (GP) kernels, where length scale and amplitude hyperparameters are sampled alongside atmospheric model parameters \citep{WangJa2020}, in order to mitigate the bias induced by correlated noise in the spectra on the model inference \citep{Greco2016}. Following the recommendations in \citet{Zhang2023}, for one iteration of the model comparison, we place a weak, truncated, normally distributed prior on mass, $M\in\mathcal{N}(\mu=1,\sigma=5, \geq0.1)\,M_{\rm J}$ based on the upper limit derived from our orbit fit and a normally distributed prior on $R\in\mathcal{N}(1.30,0.05)\,R_{\rm J}$ based on predictions from the \texttt{Sonora-Bobcat} evolutionary model \citep{Marley2021} given the system age and our mass upper limit. The posterior was sampled with 400 live points. 
\par The best-fit model from the fit including informative priors on the mass and radius has $T_\mathrm{eff}=632\pm13\,\mathrm{K}$, $\log(g)=3.7\pm0.3$, $R=1.30\pm0.03\,R_{\rm J}$, $\mathrm{[Fe/H]}=0.65\pm0.15$, $\mathrm{C/O}=0.65^{+0.05}_{-0.08}$. Figure \ref{fig:51erib_exorem_spectrum} shows the best-fitting interpolated \texttt{Exo-REM} spectrum (and the envelope of accepted solutions), compared to the observations and our $5\,\sigma$ upper limits in other NIRCam filters. 
\par As in \citet{Balmer2025}, we find that the evolutionary model informed radius prior is effectively a soft $T_\mathrm{eff}$ prior in the presence of a deterministic cloud model (that is, because there is not an explicit cloud strength parameter in the model grid, and the strength of the cloud varies depending on the other parameters, $R$ and $T_\mathrm{eff}$ are strongly correlated). In the fit without the radius or mass prior, the sampler finds a lower temperature and higher radius, with a much larger uncertainty on the surface gravity and radius, $T_\mathrm{eff}=581\pm50\,\mathrm{K}$, $\log(g)=3.75\pm0.6$, $R=1.62\pm0.31\,R_{\rm J}$.  The difference in temperature between the two model fits is relatively small given the posterior uncertainties, $\Delta T_\mathrm{eff}{=}50{\pm}30\,\mathrm{K}$, and the posterior distribution on $R$ and $T_\mathrm{eff}$ of the evolutionary model radius prior fit is fully encapsulated within the posterior distribution of the fit without this prior. Between the two fits, the abundance parameters [Fe/H] and C/O do not change.
\par With only data from the literature, the model comparison converges to effectively the same parameters as the model comparison that also includes our new F410M photometric point. The improvement when including the new JWST detection occurs where it should be expected, by partially constraining the depth of the CO$_2$ feature, the posterior on the metallicity is constrained by about $0.05\,\mathrm{dex}$ more. With a more flexible model comparison framework (for instance a retrieval with a parameterized cloud), we might expect the improvement when including the new data point to be larger.

\par 

\section{Discussion}
\label{sec:discussion}

\subsection{\hra qualitative results} \label{subsec:hr8799_discussion}

\par In \S\ref{subsec:hr8799_analysis}, we noted three interesting qualitative results from examining the NIRCam 3-5\,\um colors of the \hra planets (Figure \ref{fig:color-color}). Firstly, the planets appear distinctly offset in F430M-F410M color from the field BD population (and the field age BD companion HD~19467~B). This color traces both the wing of the CO $\nu=1-0$ absorption band head and the CO$_2$ $\nu_3$ absorption feature. At solar metallicity, this CO$_2$ feature is muted, which establishes the approximately solar metallicity field sequence in Figure \ref{fig:color-color}, traced by the blue, cloudless \texttt{Sonora-Bobcat} evolutionary and atmospheric models for 3~Gyr. To explain the offset towards bluer F430M-F410M, we over plot the corresponding 30~Myr model, but while this shifts towards blue colors, alone it doesn't appear to explain the half a magnitude difference between the \hra planets and the field. At higher metallicities, however, the 4.3\um CO$_2$ feature becomes more prominent, as evidenced by the \texttt{Sonora-Bobcat} models with $\mathrm{[M/H]}=+0.5\,\mathrm{dex}$ in orange in Figure \ref{fig:color-color}. In transiting exoplanets, this feature was hinted at by early \textit{Spitzer} photometry \citep[e.g.,][]{Wakeford2018} and recently confirmed by JWST \citep{JWSTTransitingExoplanetCommunityEarlyReleaseScienceTeam2023}, and strongly correlates with atmospheric metallicity \citep[see, e.g., Figure 13 in][]{Rustamkulov2023}. This is strong evidence that the atmospheres of the \hra planets are, at least by half a dex, enriched in metals compared to the solar value. The uncertain chemical composition of the host star, however, brings into question exactly how much of this enrichment is due to the system itself or the planets' formation. Given its $\lambda$-Boo type \citep{Gray1999}, its iron abundance of $\mathrm{[Fe/H]}=-0.5$ \citep{WangJi2020} is not a reliable tracer of its proplanetary disk abundance; it has effectively solar surface abundances for C, N, and O \citep{Sadakane2006}. Still, assuming an approximately solar metallicity, an atmospheric enrichment of even $+0.5\,\mathrm{dex}\simeq3{\times}$ solar would represent a significant fraction of solids accreted during the planet formation process for planets that have super-Jovian masses. In \citet{Marley2012} it was suggested that enriched atmospheres could improve the quality of atmospheric model fits to the \hrc and \hrd photometry, because solar metallicity models were unable to produce radii consistent with evolutionary models. In \citet{Nasedkin2024}, an exhaustive retrieval analysis of ground-based observations revealed an apparently massive enrichment in metallicity for all four planets, with values ranging from $\mathrm{[M/H]}=1-2$ for the planets. Whether such a large amount of metals could be accreted onto the planets remains to be seen; for instance the transiting giant planet HAT-P-20~b \citep{Bakos2011} has a similarly large mass ($7.2\,M_\mathrm{J}$) and high metallicity, but orbits a metal rich host star. Even if the exact metallicity of each planet is less than indicated in \citet{Nasedkin2024}, our result agrees qualitatively with theirs, namely that the \hra planets are likely metal-rich compared to the field and to their host. One subtlety in this interpretation is that, due to the vertical mixing in these atmospheres that drives chemical disequilibrium, the abundance of CO$_2$ becomes quenched higher up in the atmosphere, after CO, CH$_4$, and H$_2$O have been quenched \citep{Zahnle2016}. This effect is not necessarily captured in every modeling framework, but has for instance been explored more recently in the context of JWST observations at these wavelengths \citep{Hoch2024}, and the magnitude of this effect should affect the precise abundance determination and not the overall conclusions we present.

\par Secondly, we noted that the planets exhibit a wider spread in F430M-F410M and F460M-F410M colors than their NIR colors, spanning nearly the entire field M-L-T sequence from \hrd to \hrb in Figure \ref{fig:color-color}. This indicates that the planets' atmospheres are quite distinct, potentially owing to their formation in a radially differentiated protoplanetary disk. 
\par Thirdly, we find an apparent sequence in the outer three planets, which have increasingly bluer F430M-F410M and F460M-F410M colors. The notable exception is \hre, which has colors similar to \hrc. These two results agree qualitatively with some of the major findings of \citet{Nasedkin2024}, who found that the planets all have C/O ratios that vary with separation, decreasing from b to d, before increasing again for e. Visually, the strength of the F460M-F410M colors of each planet, which trace the 4.6\um CO feature, are correlated with the strength of the 2.3\um CO bandhead in the VLTI/GRAVITY spectra of each planet \citep[][Figure 3]{Nasedkin2024}. Core accretion with pebble drift and evaporation could favor both the apparent enriched metallicities of the planets \citep{Bitsch2023} and their radially increasing C/O ratios \citep{Schneider2021a, Schneider2021b, Molliere2022}. 

\par Any strong claim about the formation of these planets will require detailed atmospheric and protoplanetary disk modeling, since connecting observable atmospheric abundances to formation history \citep[e.g.,][]{Oberg2011} is heavily influenced by the evolution and dissipation of the disk over time \citep[e.g.,][]{Molliere2022}. We note, however, that for our third observation about the 3-5\um colors of the four planets, that there exists a sequence in 3-5\um color from \hrd to \hrb, broken only by \hre is tantalizing in the context of discussion in \citet{Molliere2022}; this sequence could be explained by formation at a given point in the disk's chemical evolution, where, past the CO iceline the C/O ratio of accreted solids remains constant and the C/O ratio of accreted gases exhibits a decreasing sequence (their Figure 3). If \hre migrated inward during its formation as suggested by \citet{Molliere2020}, that could explain its apparent departure from this sequence, as it may have crossed compositional boundaries during migration.

\subsection{\cEb Results} \label{subsec:cERi_discussion}

\par Our updated orbit fit agrees well with previous results that indicate a high-eccentricity \citep{Maire2019, DeRosa2020, Dupuy2022}. While the data does not wholly rule out low-eccentricity solutions, it strongly prefers high eccentricities $e{=}0.57_{-0.09}^{+0.03}$, with only a low-density tail of solutions extending down to $e{\sim}0$. Eccentricities above $e=0.6$ are firmly ruled out by the data. A large eccentricity at a young system age is interesting, as, assuming the planet formed within a disk that damped its initial eccentricity, it appears to indicate some dynamical event has subsequently excited the planet's eccentricity. Explanations could be planet-planet scattering \citep[e.g.,][]{Ford2008} and/or a stellar flyby \citep[e.g.,][]{Laughlin1998, Kenyon2004, Rodet2017}. Higher-precision astrometry, from, e.g., VLTI/GRAVITY, could place tighter constraints on the planet's eccentricity and other orbital elements and could even monitor the system for epicyclic motion that could be due to an unseen inner, scattering planet \citep{Lacour2021}. The system's low mid- and far-infrared luminosity \citep{Rebull2008, Riviere-Marichalar2014} could indicate the disruptive dissipation of its debris disk, unlike the preserved and relatively dynamically stable \hra system, which hosts brighter debris rings \citep{Faramaz2021, Boccaletti2023}. Alternatively, recent studies have shown that the dissipation of an eccentric protoplanetary disk itself can result in high planetary eccentricities \citep{Li2023}. The massive disk surrounding the early-type host could have been truncated by the forming planetary system, and while undergoing nonadiabatic cooling could have become eccentric. The results in \citet{Li2023} indicate that even for small disk eccentricities ($\lesssim0.05$), the resulting planetary eccentricity can become significant (0.1-0.6). 
\par We also find that the absolute astrometry from the host star results in an upper limit on the planetary mass $M_{\rm b}{<}9.0\,M_\mathrm{J}$ at $3\,\sigma$, which follows previous work \citep{DeRosa2020, Dupuy2022}, and indicates the firmly planetary nature of the companion. We used this mass upper limit to place priors on our atmospheric model comparison.
\par By placing an informed prior on the radius of the planet based on evolutionary models and our dynamical mass constraint \citep[following][]{Zhang2023}, we were able to find physically consistent atmospheric and bulk parameter solutions, though this is partially thanks to the planet's spectral type and therefore relatively lower cloud opacity compared to the \hra planets. Our best-fitting \texttt{Exo-REM} spectrum has a lower effective temperature ($T_{\rm eff}=630\,\mathrm{K}$) than found for the planet in some previous retrieval results, e.g. \citet[$T_{\rm eff}=810\,\mathrm{K}$]{Brown-Sevilla2023}, but agrees well with other RCE grid model results, \citet[$T_{\rm eff}=600{-}730\,\mathrm{K}$]{Rajan2017} and \citet[$T_{\rm eff}=680\,\mathrm{K}$]{Madurowicz2023}, and some of the retrieval permutations in \citet{Whiteford2023}. In particular, \citet{Madurowicz2023} used a custom grid of \texttt{PICASO} models that have clouds and chemical disequilibrium, like our model of choice, and demonstrated that these attributes are required to match the \textit{L'}  and \textit{M} band photometry of the planet. Our best-fitting model also has an enriched metallicity, $\mathrm{[M/H]}=0.65$ and moderately enriched C/O compared to solar, $\mathrm{C/O}=0.65$. This indicates an atmospheric metallicity enriched by about $5\times$ compared to the stellar value \citep[$\mathrm{[Fe/H]}=-0.1$,][]{Koleva2012, Rajan2017}.

\section{Conclusions}
\label{sec:conclusion}

\par JWST's impressive wavefront stability and pointing accuracy continues to enable excellent high-contrast observations. In this paper we presented some of the first observations using the shared-risk ``narrow" offset position on the NIRCam \texttt{LWB} coronagraph, demonstrating the performance of the coronagraph at ${<}4\lambda/D$ by detecting the close-in planets \cEb (0\farcs28) and \hre (0\farcs4). Our contrast performance at close separations (${<}0\farcs5$) using the ``narrow" offset position improves on previous performance using the filter-dependent offsets for the \texttt{LWB} and the round masks, and opens the door for characterization studies of known, closely separated giant planets with JWST (for instance, a survey of the CO$_2$/CO absorption or the bolometric luminosity of known imaged companions). 
\par The 3-5\um photometry of the HR~8799 system reveal a variety of interesting results, namely, stronger CO$_2$ absorption than expected for solar abundance atmospheres, implying enriched metallicities, and a significant differentiation in the compositions of the planets that extends radially outward from \hrd to \hrb. These qualitative results can be interpreted as evidence in favor of some recently proposed core accretion formation models, as was suggested in \citet{Nasedkin2024}, but necessitates more detailed modeling to truly validate. Our results present an important constraint on the temperature structure of these planet's atmospheres, in the context of upcoming studies that will measure the continuum subtracted spectra of these planets with moderate-resolution spectroscopy \citep[GTO 1188, following][]{Ruffio2024}. As in high-resolution stellar atmospheric modeling, temperature and surface gravity degeneracies often need to be addressed with strong priors based on photometric data. These degeneracies are even more important to account for in substellar atmospheric analyses at moderate or high resolution when cloud opacities are significant, but cloud information content is not always preserved in the data \citep{Xuan2022}. Future work will require the simultaneous fitting of photometric/low-resolution data with moderate- and high-resolution data to accurately assess the abundances of these planets \citep[as in, e.g.][]{Wang2023}.
\par Our detection of \cEb in the mid-infrared allows us to place an updated constraint on the planet's atmospheric properties, which are key to understanding its formation history. The updated orbit fit continues to prefer high-eccentricity solutions ($e{=}0.57_{-0.09}^{+0.03}$) and low secondary masses ($M_{\rm b}{<}9.0\,M_\mathrm{J}$ at $3\,\sigma$), which could indicate that some dynamical process has occurred early in this young system's history. This could hint at scattering from an unseen inner or ejected planet, a recent stellar flyby, or a disk-driven process. Like the \hra planets, the extraction and interpretation of upcoming spectroscopic observations of the planet with JWST (GO 3522) will benefit from the photometry and upper limits we have measured here.
\par Additionally, the success of targeted, accelerating star studies \citep[e.g.,][]{DeRosa2023, Franson2023, Mesa2023}, where the position of the companion can be predicted with confidence \citep[e.g., Figure 1 in][]{Rickman2022} opens the door for targeted, closely separated planet searches with this mode. If a companion's location can be predicted to better than $\pm45^\circ$ on sky and it would be undetectable from the ground at shorter wavelengths but visible from 3-5\um, JWST/NIRCam might be uniquely able to observe it using the \texttt{LWB}/narrow offset.

% \begin{acknowledgments}
\section*{Acknowledgments}
% \par \todo{acknowledge wopr, etc?}
\par We thank the anonymous reviewer for their positive and constructive report. We also thank the program coordinators for GTO 1194 and 1412, Crystal Mannfolk and Beth Perriello, respectively, for their assistance in the  completion of our observations. W.O.B. thanks Amanda Chavez and Jason Wang for making their orbit fits to the HR~8799 planets available on \texttt{whereistheplanet} and for consulting with us on their accuracy, Evert Nasedkin for conversations regarding atmospheric retrievals, and Aarynn Carter for sharing the NIRCam \texttt{335R} F410M contrast curve for comparison in Figure \ref{fig:f410m_contrast_curve_hr8799}.
\par This paper reports work carried out in the context of the \emph{JWST} Telescope Scientist Team (PI M.~Mountain). Funding is provided to the team by NASA through grant 80NSSC20K0586. Based on observations with the NASA/ESA/CSA \emph{JWST}, obtained at the Space Telescope Science Institute, which is operated by AURA, Inc., under NASA contract NAS 5-03127. The data presented in this article were associated with GTO 1194 (PI: C.~Beichman) and 1412 (PI: M.~Perrin) and obtained from the Mikulski Archive for Space Telescopes (MAST) at the Space Telescope Science Institute. This data can be accessed via \dataset[DOI 10.17909/e4x8-ft26]{http://dx.doi.org/10.17909/e4x8-ft26}.
\par We are grateful for support from NASA through the JWST/NIRCam project, contract number NAS5-02105 (PI: M. Rieke, University of Arizona), and to M. Rieke for the generous invitation to attend and present these results at the NIRCam Team Meeting on March 2nd, 2025.
\par Part of this work was carried out by W.O.B. at the Advanced Research Computing at Hopkins (ARCH) core facility (rockfish.jhu.edu), which is supported by the National Science Foundation (NSF) grant number OAC1920103.
\par This research has made use of the SVO Filter Profile Service, funded by MCIN/AEI/10.13039/501100011033/ through grant PID2020-112949GB-I00
\par This research has made use of the VizieR catalog access tool, CDS, Strasbourg, France (DOI: 10.26093/cds/vizier).
\par This work has made use of data from the European Space Agency (ESA) mission {\it Gaia} (\url{https://www.cosmos.esa.int/gaia}), processed by the {\it Gaia} Data Processing and Analysis Consortium (DPAC, \url{https://www.cosmos.esa.int/web/gaia/dpac/consortium}). Funding for the DPAC has been provided by national institutions, in particular the institutions participating in the {\it Gaia} Multilateral Agreement.
\par This publication makes use of data products from the Two Micron All Sky Survey, which is a joint project of the University of Massachusetts and the Infrared Processing and Analysis Center/California Institute of Technology, funded by the National Aeronautics and Space Administration and the National Science Foundation. 
\par W.O.B. acknowledges their cat, Morgoth, for her ``encouragement."
% \end{acknowledgments}

% \par In this paper we:

% \begin{enumerate}
%     \item Took a bad picture
%     \item Blew up a star with math
%     \item Consulted the bones
%     \item Found that not every planet is the same (who would have guessed?)
% \end{enumerate}

\appendix

\section{Filter Gallery} \label{appendix:imaging}

\begin{figure}

\gridline{\fig{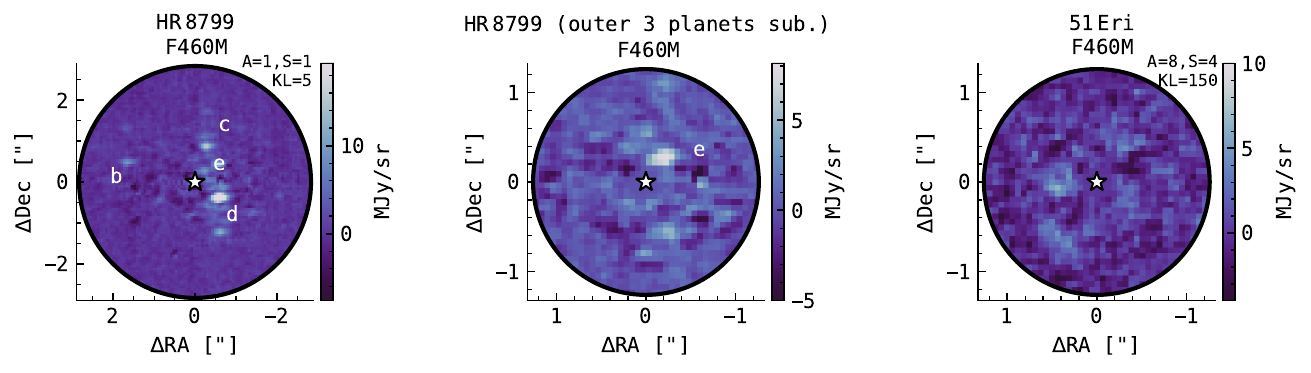}{\textwidth}{}}

\gridline{\fig{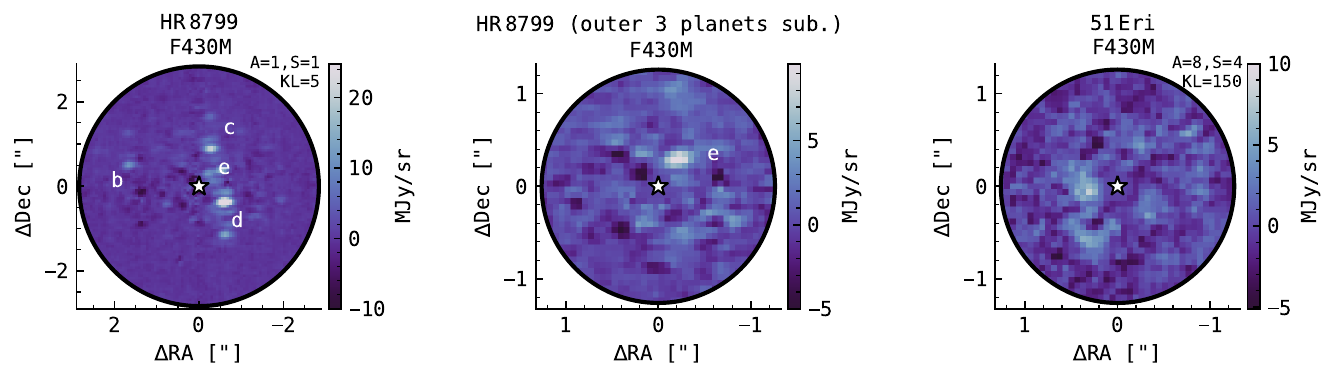}{\textwidth}{}}

\gridline{\fig{figures/imaging_summary/NIRCam_Bar_NARROW_F410M_summary_altcolor.pdf}{\textwidth}{}}

\gridline{\fig{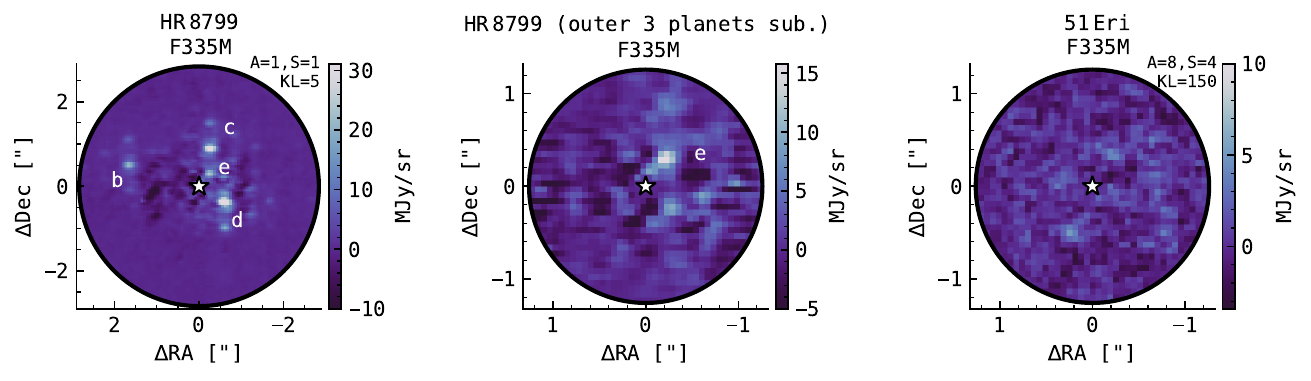}{\textwidth}{}}

\caption{Post starlight subtraction image library by filter, for F460M through F335M} \label{fig:img_summary_all_1}

\end{figure}

\begin{figure}

\gridline{
\fig{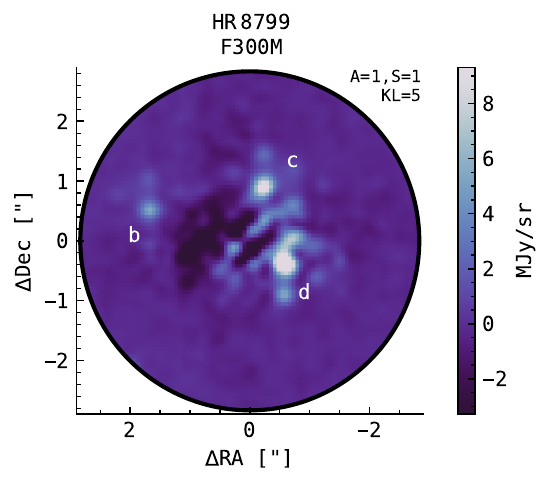}{0.4\textwidth}{}
\fig{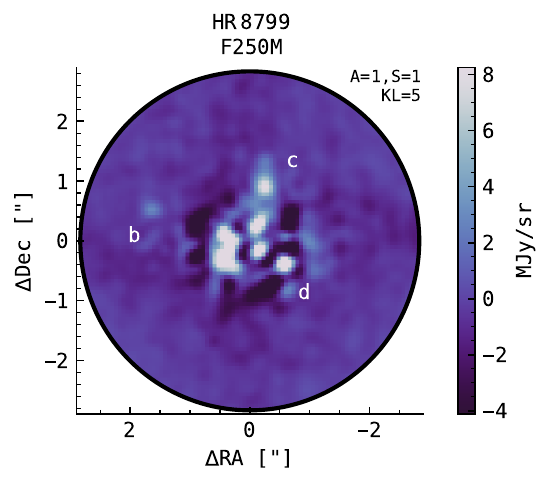}{0.4\textwidth}{}
}

\gridline{
\fig{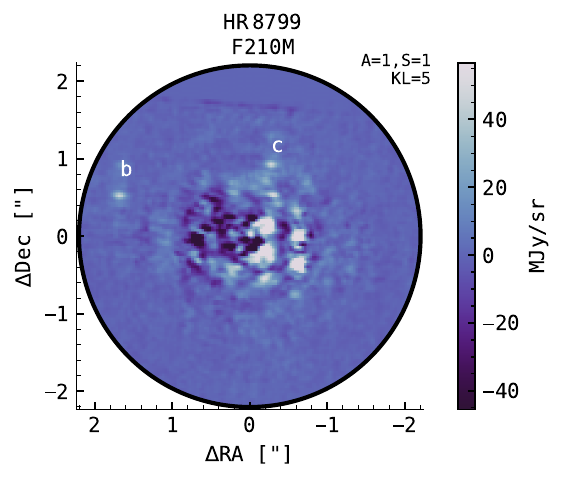}{0.4
\textwidth}{}
\fig{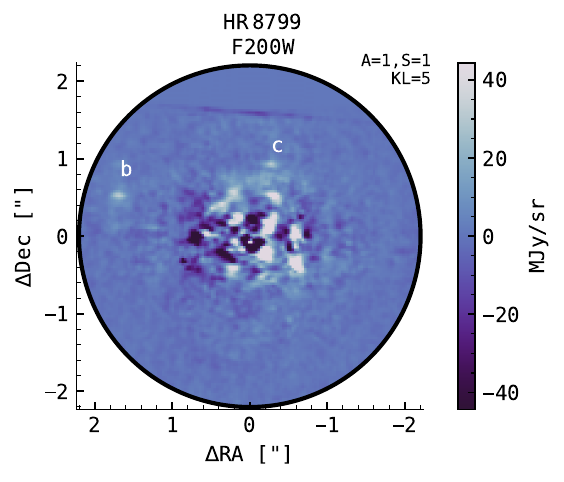}{0.4
\textwidth}{}
}

\gridline{
\fig{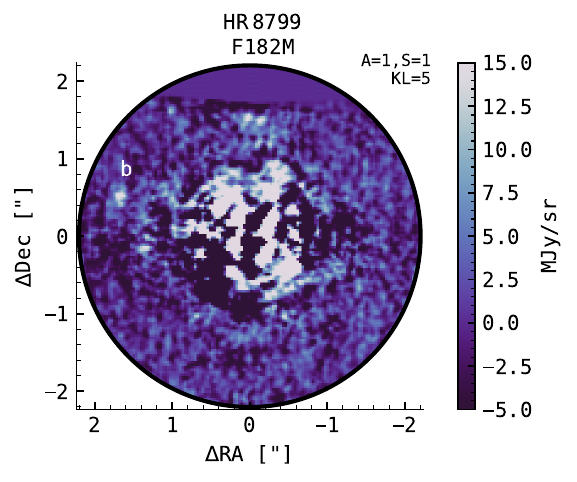}{0.4\textwidth}{}
}

\caption{Post starlight subtraction image library by filter, for F300M, F250M (LW detector) and F210M, F200W, and F182M (SW detector). ${>}5\,\sigma$ detections are labeled according to Table \ref{tab:bka}.} \label{fig:img_summary_all_2}

\end{figure}

\begin{figure}

\gridline{\fig{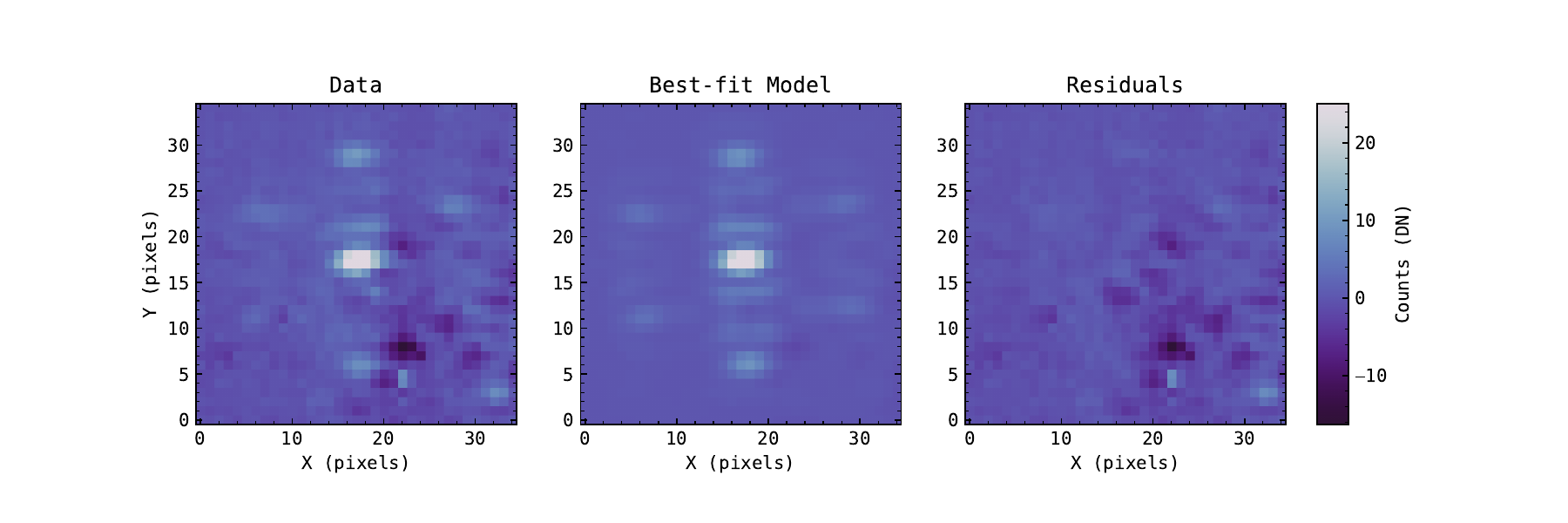}{0.6\textwidth}{\hrb}}

\gridline{\fig{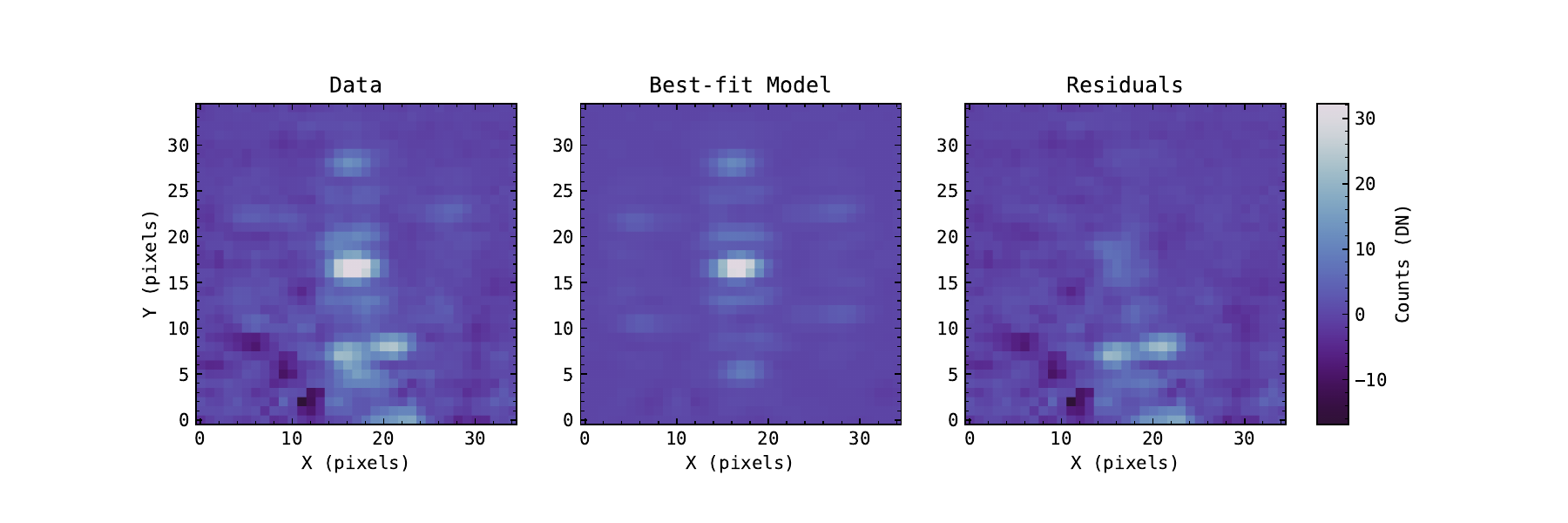}{0.6\textwidth}{\hrc}}

\gridline{\fig{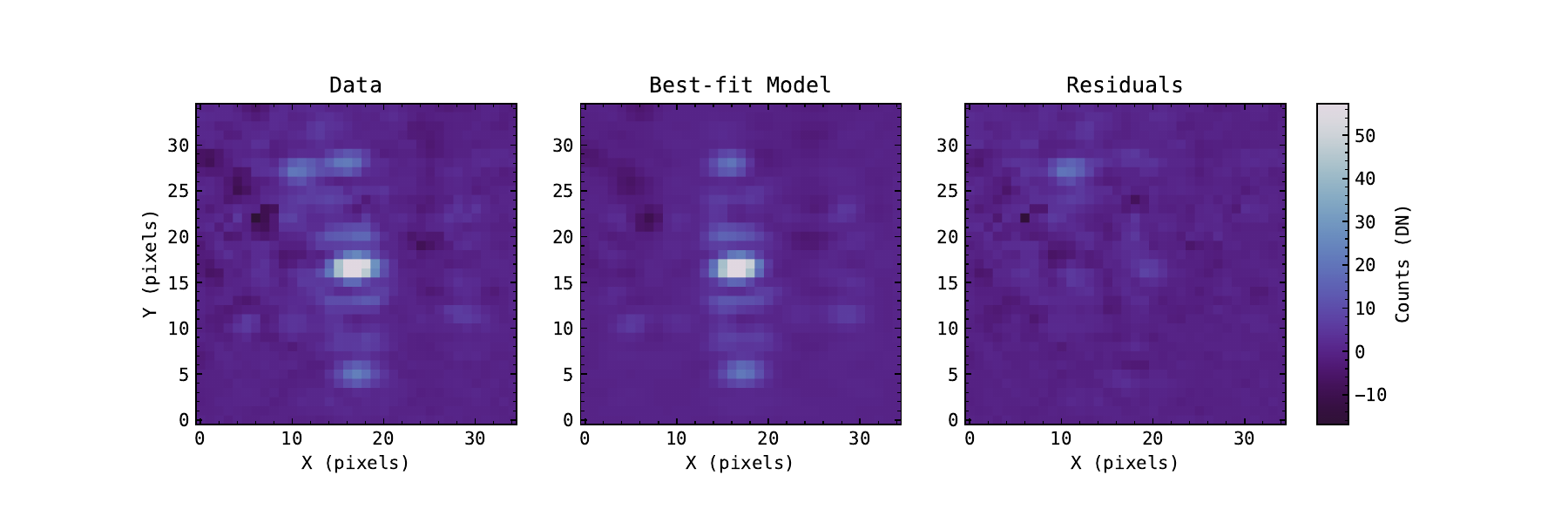}{0.6\textwidth}{\hrd}}

\gridline{\fig{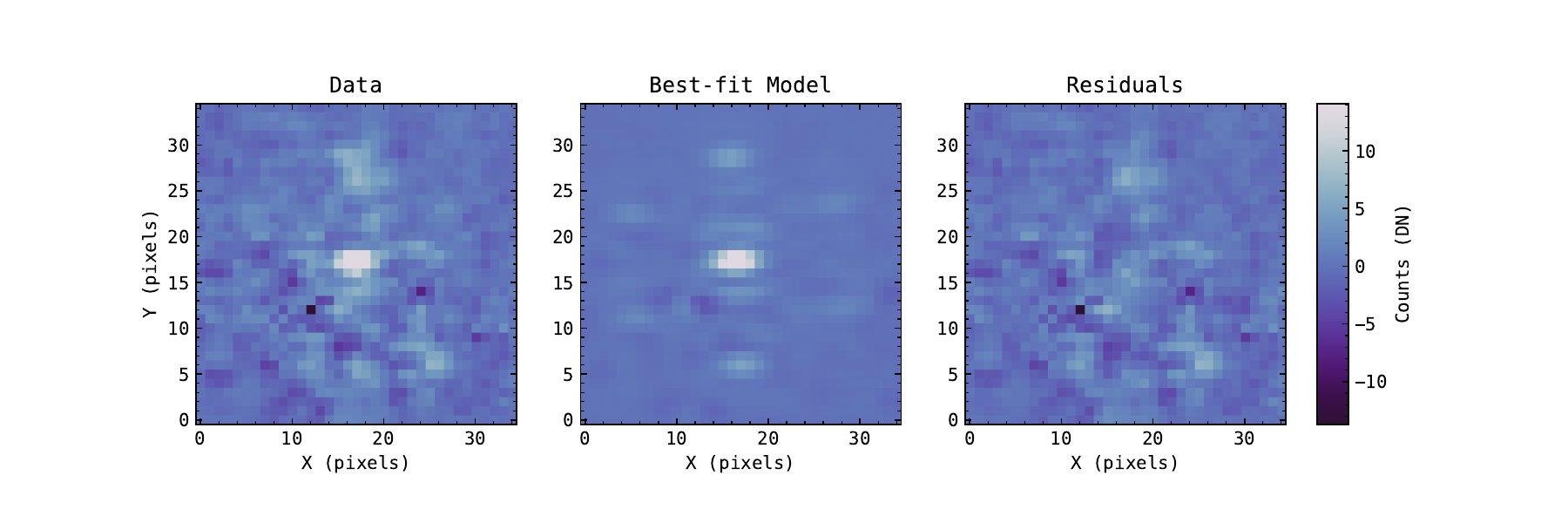}{0.6\textwidth}{\hre}}

\gridline{\fig{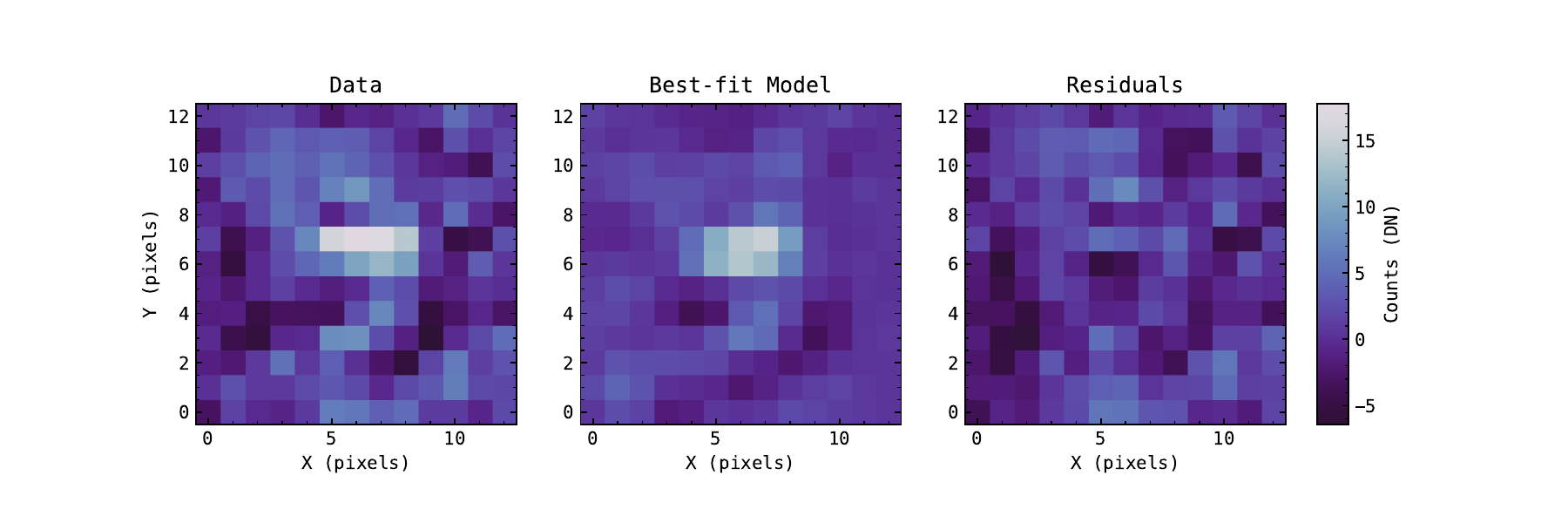}{0.6\textwidth}{\cEb}}

\caption{Example Bayesian KLIP-FM Astrometry (BKA) results for the F410M filter, corresponding to the photometry and astrometry recorded in Table \ref{tab:bka}. %\todo{In the html version this will hopefully be an interactive gallery so that all filters can be inspected}.
} 
\label{fig:bka}

\end{figure}

\begin{figure}
    \centering
    \includegraphics[width=\linewidth]{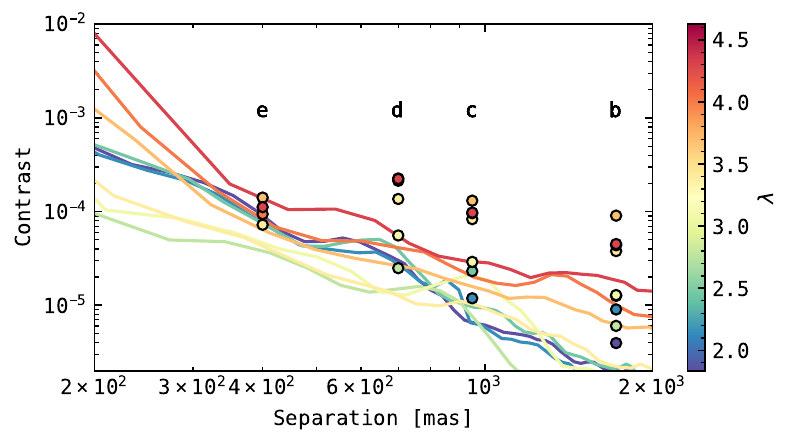}
    \caption{Following Figure \ref{fig:f410m_contrast_curve_hr8799}, calibrated $5\,\sigma$ contrast curves and measured contrasts for each filter across the \hra observations.}
    \label{fig:all_ccs}
\end{figure}

This Appendix hosts two figures that contain the starlight-subtracted images for all filters, Figures \ref{fig:img_summary_all_1} and \ref{fig:img_summary_all_2}, across both planetary systems, including images with the outer three \hra planets modeled and subtracted. There is also included a library, Figure \ref{fig:bka}, of our BKA process (data, model, and residuals) for each planet in each filter, corresponding to Table \ref{tab:bka}. Also included is the computed contrast curves (Figure \ref{fig:all_ccs}) for the \texttt{ADI+RDI} subtraction (and subsequent planet forward model subtraction) for the \hra observations. 

\section{Spatially Dependent Coronagraphic PSF Deconvolution} \label{appendix:deconv}

\par To better illustrate the strength of the detection of \hre in the presence of the outer three planets and to improve the interpretability of the images, we used \texttt{webbpsf} to deconvolve our F410M, F430M, and F460M images of the \hra system. These images were provided to the STScI Office of Public Outreach for the construction of their press release, three-color image. Beginning from the PSF-subtracted image for each roll, we perform deconvolution using a variant of the iterative Richardson-Lucy \citep[R-L;][]{Richardson1972, Lucy1974} algorithm that has been modified to accommodate (a) the shift-variant NIRCam coronagraphic PSF, and (b) the transmission of the coronagraph. This algorithm will be detailed in an upcoming study (Lawson et al. in prep.), but is summarized briefly hereafter. 

\par A compelling property of conventional R-L deconvolution is that, if it converges, it converges to the maximum likelihood solution \citep{Shepp1982}. Consider an observed 2-D image array, $\mathbf{I}$, and an operator that applies instrumental blurring to a 2-D image array, $\mathcal{B}$. For R-L deconvolution, the $(i+1)^{th}$ iteration, $\mathbf{R}_{i+1}$, is computed by multiplying the $i^{th}$ iteration, $\mathbf{R}_{i}$, by a correction term. For conventional R-L deconvolution, this correction is computed as $\mathcal{B}\left[\mathbf{I} / \mathcal{B}(\mathbf{R}_{i})\right]$, i.e., the result of blurring the quotient between the observed image and the re-blurred $i^{th}$ iteration. For a noiseless image with a shift-invariant PSF, if the $i^{th}$ iteration is exactly the true unblurred image, then the quotient is unity, and blurring the quotient results in a correction term of unity: the $(i+1)^{th}$ iteration equals the $i^{th}$ iteration. For a shift-variant PSF, blurring a uniform image yields a nonuniform result (e.g., being dimmer wherever the PSF is flatter). With the conventional R-L procedure, this effect will be present in both the numerator and denominator of the quotient and so cancels out. However, blurring that quotient produces a nonuniform correction term because of the shift-variant PSF, leading to a non-unity correction term for the true unblurred image. To recover the convergence property, we simply divide the correction term at each iteration by the result of blurring an image of ones. To account for the effect of the bar mask, the $i^{th}$ iteration is multiplied by a map of the bar mask's transmission, $\mathbf{T}$, prior to re-blurring. Denoting an image of ones as $\mathbb{J}$, our iterative correction term is thus $\mathcal{B}\left[\mathbf{I} / \mathcal{B}(\mathbf{T} \cdot \mathbf{R}_{i})\right] / \mathcal{B}(\mathbb{J})$, where all division and multiplication are performed element-wise.

\begin{figure}
\gridline{\fig{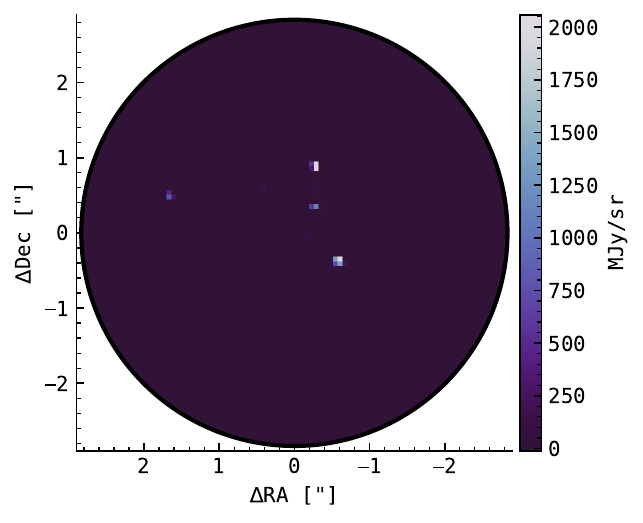}{0.48\textwidth}{}
\fig{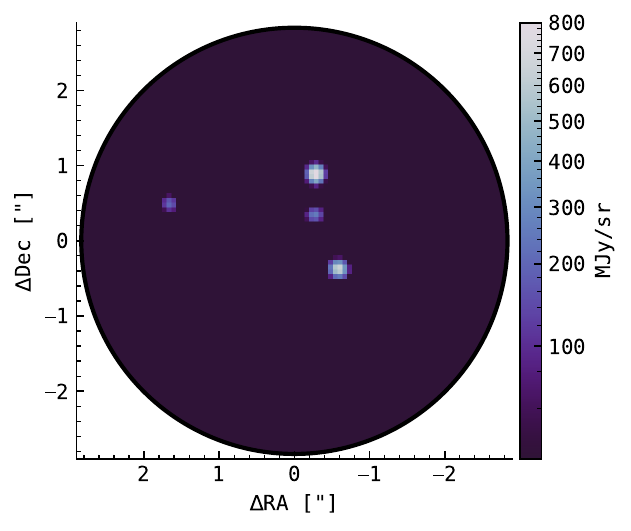}{0.48\textwidth}{}}
% twocolumn
% \gridline{\fig{figures/HR8799_NIRCam_F410M_deconvolved_altcolor.pdf}{0.4\textwidth}{}}
% \gridline{
% \fig{figures/HR8799_NIRCam_F410M_deconvolved_smoothed_altcolor.pdf}{0.4\textwidth}{}}
\caption{Deconvolution of the F410M image of \hra (Fig.\,\ref{fig:imaging_summary}) using \texttt{webbpsf} model PSFs, as described in Appendix \ref{appendix:deconv}. \textit{Left}: Deconvolved result, \textit{Right}: Top, smoothed by a Gaussian with a FWHM of 1 pixel. The deconvolution, which has accounted for the spatially dependent off-axis coronagraphic PSF as well as the coronagraph transmission function captures the intrinsic luminosity difference between the four planets in this filter (see also the calibrated photometry in Table \ref{tab:bka}). } \label{fig:deconvolved}
\end{figure}

In each case, blurring is carried out following the general procedure of \citet{Lawson2023}, modified for use of the bar mask. Here, a grid of synthetic PSFs is sampled along the bar mask using \texttt{webbpsf}. For each of 11 linearly spaced positions along the length of the bar mask, 21 samples are drawn: one at the middle of the bar plus 10 logarithmically spaced points perpendicular to the bar in both directions (231 samples in total). Each PSF is normalized such that an infinite aperture at the exit pupil would measure a total flux of one. To blur a given image, each pixel is convolved with the nearest sampled PSF. The modified R-L technique is carried out for 250 iterations for each roll, after which the rolls are combined to produce a final deconvolved image (Figure \ref{fig:deconvolved}).

\section{Retrieval Comparison} \label{appendix:retrieval_comparison}

\begin{figure*}
    \centering
    \gridline{\fig{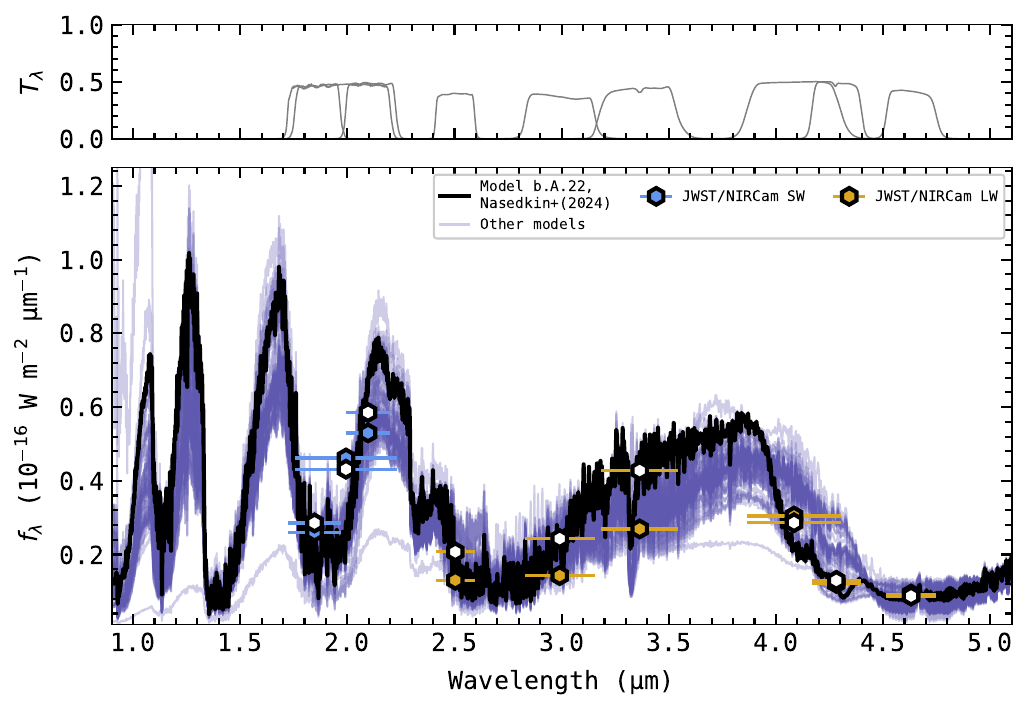
    % \gridline{\fig{figures/retrieval_comparison/HR8799b_nasedkin24_compare_alldata.pdf
    }{0.48\textwidth}{\hrb} 
    \fig{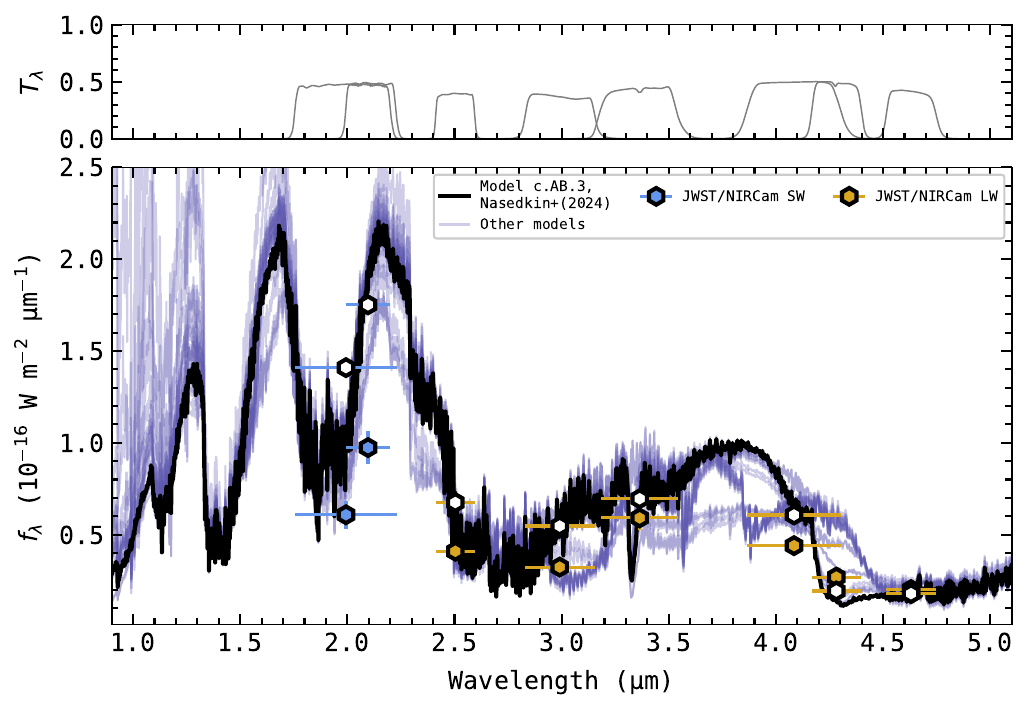
    % \fig{figures/retrieval_comparison/HR8799c_nasedkin24_compare_alldata.pdf
    }{0.48\textwidth}{\hrc} 
    }
    \gridline{
    \fig{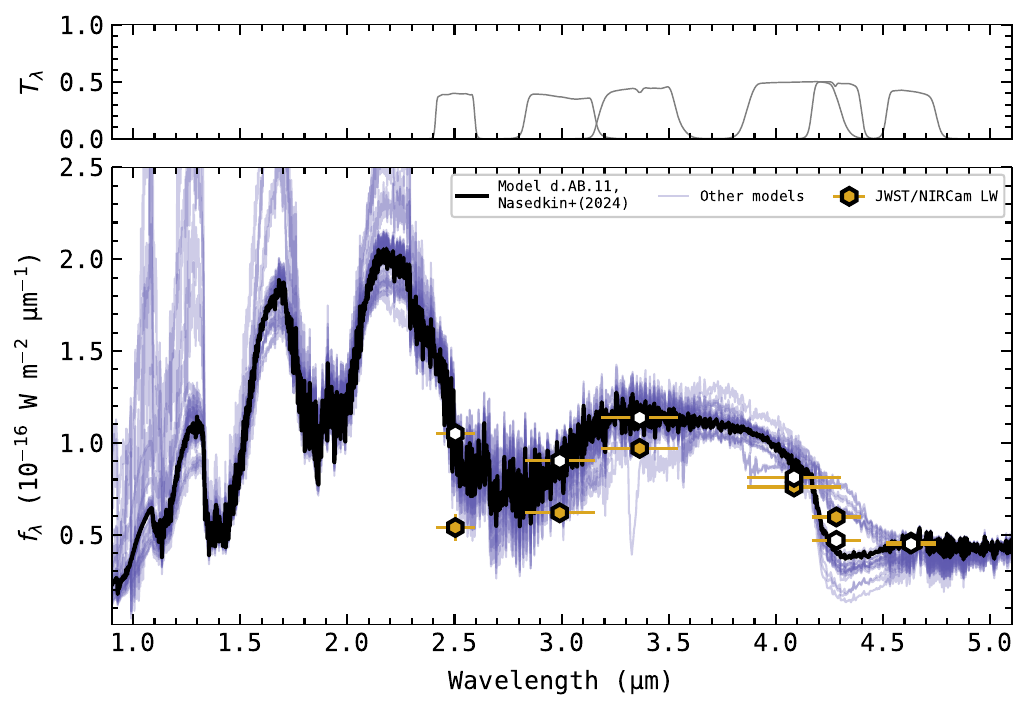
    % \fig{figures/retrieval_comparison/HR8799d_nasedkin24_compare_alldata.pdf
    }{0.48\textwidth}{\hrd}
    \fig{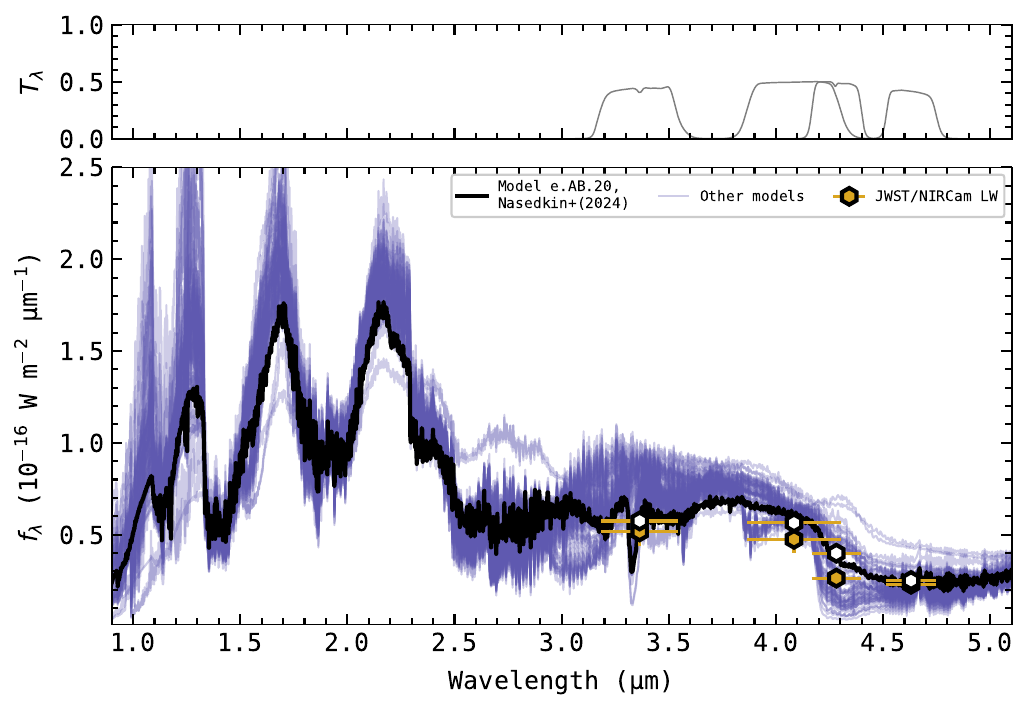
    % \fig{figures/retrieval_comparison/HR8799e_nasedkin24_compare_alldata.pdf
    }{0.48\textwidth}{\hre}
    }
    
    \caption{\texttt{LWB}/narrow observations of all four \hra planets (SW filters as blue hexagons, LW filters as gold hexagons) compared to \texttt{petitRADTRANS} models that were fit to previously available observations, primarily ground-based spectroscopy and photometry (shown in Figure \ref{fig:compare_datasets}), from \citet{Nasedkin2024}. A random model has been chosen to highlight in black, and synthesized photometry from this model for each NIRCam filter is shown as an open hexagon; all other models from the exhaustive retrieval analysis are plotted in light purple. Some of the retrieval models are consistent with some of the NIRCam photometry, but many are not, especially at shorter wavelengths near the \textit{K} band. This could be due to the absolute flux calibration of the ground-based spectra; see discussion in \S\ref{subsec:hr8799_analysis}.}
    \label{fig:all_compare_retrievals}
\end{figure*}

This Appendix hosts Figure \ref{fig:all_compare_retrievals}, which compares the measured NIRCam photometry for the \hra planets to the atmospheric retrievals fit to ground-based data from \citet{Nasedkin2024}.

\section{Orbit Posterior Distribution} \label{appendix:corner_plots}

\begin{figure}
    \centering
    \includegraphics[width=\textwidth]{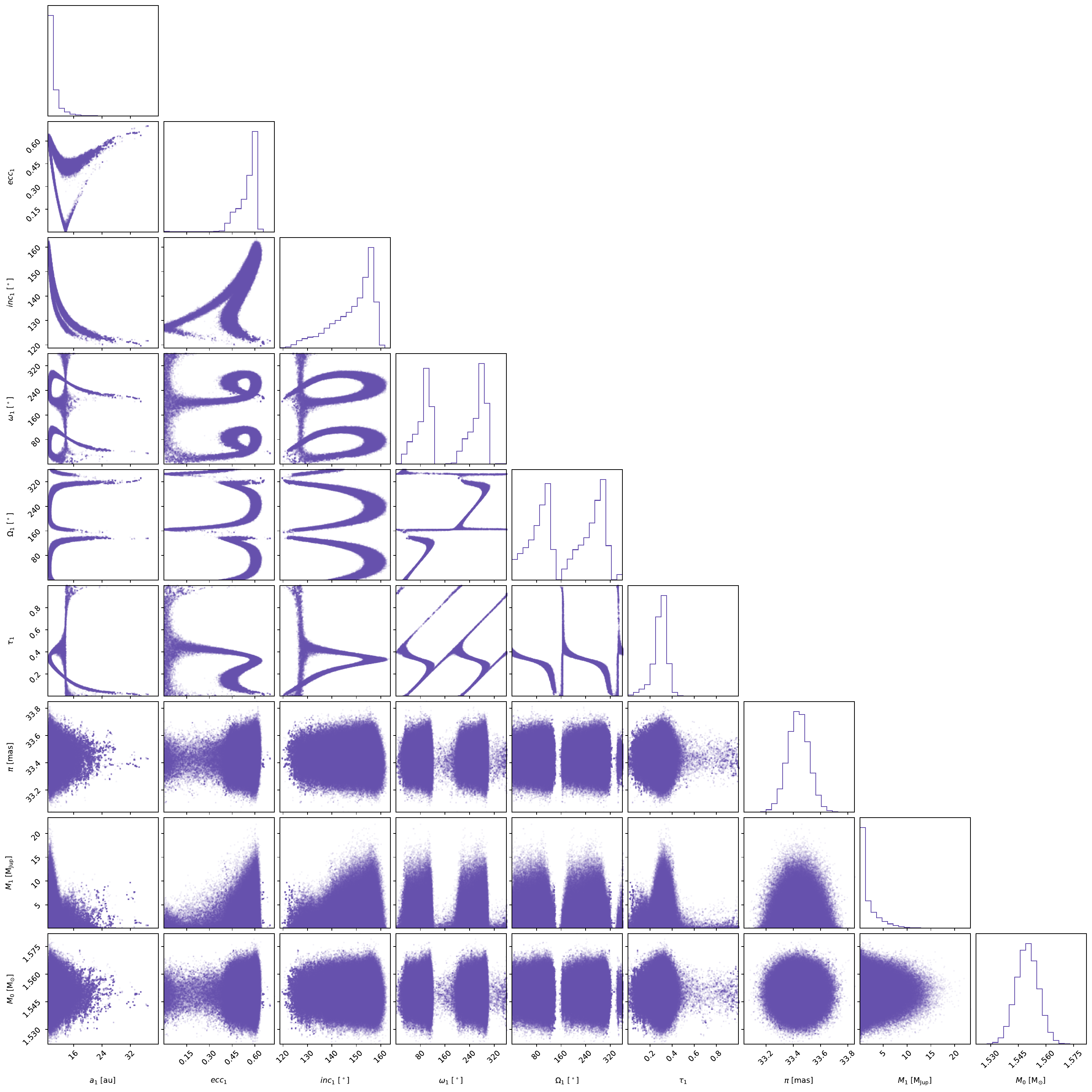}
    % KL: "This is the most haunted corner plot I've ever seen. "
    % LW: "It looks like it was directed by Tim Burton."
    \caption{Visual orbit parameters and dynamical masses of the \cE system, from our orbital analysis (\S\ref{subsec:51eri_analysis}) using \texttt{orbitize!}. Parameters with respect to \cEb are denoted with subscript ``1," and parameters with respect to \cE~A with subscript ``0." Complex correlations between the parameters describing the visual orbit are apparent, as in previous orbit fits \citep{Maire2019, DeRosa2020, Dupuy2022}; the updated measurement from JWST/NIRCam confirms the preference for high-eccentricity solutions ($e=0.58$) and a moderate semi-major axis ($a=9.34\,\mathrm{au}$) that was found in those works. The inclusion of absolute astrometry from the HGCA provides an upper limit on the planet's mass, $M_{\rm b}{<}9.0\,M_\mathrm{J}$ at $3\,\sigma$.}
    \label{fig:51eri_orbit_corner}
\end{figure}

This Appendix hosts Figure \ref{fig:51eri_orbit_corner}, which shows the posterior of orbit fits from our \texttt{orbitize!} run described in \S\ref{subsec:51eri_analysis} and visualized in Figure \ref{fig:51eri_orbit}.

\newpage
\bibliography{hr8799bar}{}
\bibliographystyle{yahapj}  % avoids the "----" style, which is nice but makes it more difficult for e.g. NASA ADS to make links

\end{document}